\documentclass[11pt,a4paper,leqno]{article}
\usepackage[utf8]{inputenc}
\usepackage{authblk}
\usepackage{geometry}
\usepackage{xcolor}
\usepackage[round]{natbib}
\usepackage{graphicx}
\usepackage{pdflscape}  
\usepackage{appendix}
\usepackage{threeparttable}
\usepackage{booktabs}

\usepackage{amssymb}
\usepackage{amsbsy}
\usepackage{bm}
\usepackage{amsmath}
\usepackage{amsthm}
\usepackage{graphicx}
\usepackage{mathtools}
\usepackage{rotating}
\usepackage{setspace}
\usepackage{footmisc}
\usepackage{color, colortbl}
\definecolor{ashgrey}{rgb}{0.7, 0.75, 0.71}
\definecolor{columbiablue}{rgb}{0.61, 0.87, 1.0}
\definecolor{coral}{rgb}{1.0, 0.5, 0.31}

\DeclareMathOperator*{\argmin}{arg\,min}

 \newcommand{\A}[1]{\cellcolor{gray}}
 \newcommand{\B}[1]{\cellcolor{lightgray}}

\usepackage{enumitem}
\newlist{steps}{enumerate}{1}
\setlist[steps,1]{label = Step \arabic*:}

\usepackage{adjustbox}
\usepackage{dcolumn}
\newcolumntype{d}[1]{D..{#1}} 

\usepackage{caption}
\usepackage{subcaption}
\definecolor{nblue}{HTML}{000660}
\usepackage[colorlinks=true,urlcolor=nblue,linkcolor=nblue,citecolor=nblue]{hyperref}

\newcommand*{\myeqref}[2][Eq.~]{%
  \hyperref[{#2}]{#1(\ref*{#2})}%
}
\def\equationautorefname#1#2\null{%
  Eq.#1(#2\null)%
}

\begin{document}
\title{\textbf{Nonlinearities in Macroeconomic Tail Risk through the Lens of Big Data Quantile Regressions}}
\author{}
\date{}

\maketitle
\thispagestyle{empty}
\vspace*{-6.5em} 
\begin{center}
\end{center}

\normalsize

\vspace*{-0.75em}
\begin{minipage}{.49\textwidth}
  \large\centering Jan \textsc{Pr\"user}\\[0.25em]
  \small TU Dortmund\\ Department of Statistics
\end{minipage}
\begin{minipage}{.49\textwidth}
  \large\centering Florian \textsc{Huber}\footnotemark{}\\[0.25em]
  \small University of Salzburg\\ Department of Economics
\end{minipage}

\footnotetext{
We would like to thank the editor, Mike McCracken, three anonymous referees, and  participants at the International Symposium on Forecasting 2022 at the University of Oxford and at the Statistische Woche in M\"unster for helpful comments. Jan Pr\"user gratefully
acknowledges the support of the German Research Foundation (DFG, 468814087). Please address correspondence to: Jan Prüser. Department of Statistics, TU Dortmund. \textit{Address}: CDI Building, Room 122,
44221 Dortmund, Germany. \textit{Email}: \href{mailto:prueser@statistik.tu-dortmund.de}{prueser@statistik.tu-dortmund.de}.}
\vspace*{2em}

\doublespacing
\begin{center}
\begin{minipage}{0.9\textwidth}
\noindent\small \textbf{Abstract.} Modeling and predicting extreme movements in GDP is notoriously difficult and the selection of appropriate covariates and/or possible forms of nonlinearities are key in obtaining precise forecasts.  In this paper, our focus is on using large datasets  in  quantile regression models to forecast the conditional distribution of US GDP growth. To capture possible non-linearities, we include several nonlinear specifications. The resulting models will be huge dimensional and we thus rely on a set of shrinkage priors. Since Markov Chain Monte Carlo estimation becomes slow in these dimensions, we rely on fast variational Bayes approximations to the posterior distribution of the coefficients and the latent states. We find that our proposed set of models produces precise forecasts. These gains are especially pronounced in the tails. Using Gaussian processes to approximate the nonlinear component of the model further improves the good performance, in particular in the right tail.\\\\ 

\end{minipage}
\end{center}

\normalsize\newpage\renewcommand{\footnotelayout}{\setstretch{2}}

\section{Introduction}
Modeling and predicting the conditional distribution of output growth has attracted considerable academic attention in recent years. Starting at least with the influential paper by \cite{adrian2019vulnerable}, focus has shifted towards analyzing whether there exist asymmetries between a predictor (in their case financial conditions) and output growth across different quantiles of the empirical distribution. Several other papers \citep{adrian2018term, Ferrara2019, gonzalez2019growth, delle2020modeling, plagborg2020growth, reichlin2020financial, figueres2020vulnerable,adams2021forecasting,mitchell2022nowcasting} have started to focus on modeling full predictive distributions using different approaches and information sets. However, most of these contributions have been confined to models which exploit small datasets and, at least conditional on the quantile analyzed, assume linear relations between GDP growth and the predictors. 


Times of economic stress such as the global financial crisis (GFC) or the Covid-19 pandemic have highlighted that exploiting information contained in many time series and allowing for nonlinearities improves predictive performance  in turbulent periods \citep[see, e.g.,][]{huber2020nowcasting}. Since economic dynamics change in volatile economic regimes, models that control for structural breaks allow for different effects of economic shocks over time or imply nonlinear relations between GDP growth and its predictors often excel in forecasting applications \citep[see][]{d2013macroeconomic, carriero2016common, adrian2021multimodality, chkmp2021tail,pfarrhofer2022modeling, huber2020nowcasting}. Moreover,  another important empirical regularity is that the set of predictors might change over time. This is because variables which are seemingly unimportant in normal periods (such as financial conditions) play an important role in recessions and yield important information on future behavior of output growth.


This  discussion highlights that the effect of predictors on output growth depends on the quantile under consideration and thus appears to be state dependent and modeling the transition might call for nonlinear econometric models.  The key challenge, however, is to identify the different determinants of GDP growth across quantiles while taking possible nonlinearities into account.  In this paper, we aim to solve these issues by proposing a Bayesian quantile regression (QR) which can be applied to huge information sets, and which is capable of capturing nonlinearities of unknown form. Our model is a standard QR model that consists of two parts. The first assumes a linear relationship between the covariates and quantile-specific GDP growth, whereas the second component assumes an unknown and possibly highly nonlinear relationship between the two. The precise form of nonlinearities is captured through three specifications. One is parametric and based on including polynomials up to a certain order, whereas the remaining two are nonparametric. Among these nonparametric specifications we include B-splines \citep[see][]{shin2020functional} and Gaussian processes \citep[see][]{williams2006gaussian}. Both have been shown to work well when it comes to function estimation and forecasting.

The combination of a linear and nonlinear specification implies that the dimension of the parameter space increases substantially. Since all these models can be cast in terms of a linear regression conditional on appropriately transformed covariates, we can use regularization techniques to decide on whether more flexibility is necessary and which variables should enter the model. We achieve this through several popular shrinkage priors that have excellent empirical properties in large dimensions and are relatively easy to implement. These shrinkage priors enable us to select promising subsets of predictors and the degree of nonlinearities for each quantile separately.  

Posterior inference using Markov Chain Monte Carlo (MCMC) techniques in these dimensions proves to be an issue because we have to estimate a large-scale regression model for all quantiles of interest. This procedure needs to be repeated a large number of times if we wish to carry out an out-of-sample forecasting exercise. To reduce the computational burden enormously we estimate the QRs using Variational Bayes (VB).\footnote{For an introduction, see \cite{blei2017variational} and an algorithm for QRs is provided in \cite{lim2020sparse}.} This estimation strategy approximates the exact full conditional posterior distributions with simpler approximating distributions. These approximating densities are obtained by minimizing the Kullback-Leibler (KL) distance between some known density $q$ and the exact posterior distribution $p$. Hence, integration in huge dimensions is replaced by a simpler optimization routine. Our approach is  fast and allows for computing all results of our forecasting exercise without the use of high performance computing environments. As compared to, e.g.,  \cite{kohns2021decoupling} and \cite{mitchell2022nowcasting}, who estimate large QRs  under flexible Bayesian shrinkage priors, our techniques are highly scalable and can be applied to models even larger than the ones we consider in this paper.

We apply our techniques to the large dimensional FRED-QD dataset  \citep{mccrackenngdata} and focus on single and multi-step-ahead forecasting  of US GDP growth over a hold-out period ranging from 1991Q2 to {2021Q3}.  The different nonlinear models we consider are high dimensional and feature up to over 1,000 coefficients per equation.  

The empirical results can be summarized as follows. We show that forecasts from models estimated with VB are very similar to the ones obtained from estimating the different models using MCMC. When we consider the forecasts from the large-scale VB QRs we find that   
using huge information sets and nonlinear models in combination with priors that introduce substantial shrinkage pays off for tail forecasts. In both tails, forecast improvements relative to a large-scale QR estimated using MCMC techniques and a Minnesota prior are sizable. When we focus on the center of the distribution, the differences become smaller. Once we allow for nonlinearities, we find some improvements in predictive accuracy over the full hold-out period. These improvements are mostly related to the right tail while in the left tail accuracy gains are smaller. Comparing the different nonlinear specifications reveals that Gaussian processes offer the largest improvements vis-\'{a}-vis the linear QR. Once we condition on the more volatile second half of the sample (starting in $2006$Q2), accuracy premia obtained from nonlinear models increase substantially. In particular, nonlinear models yield appreciable improvements in tail forecasting accuracy during the GFC and the pandemic. This indicates that a successful tail forecasting model should be able to extract important information from  huge datasets, while controlling for possibly nonlinear relations during turbulent periods. When we focus on the key properties of the proposed priors, we observe that priors that imply a dense model (characterized by many small coefficients) yield good tail forecasts. 

The paper is structured as follows. The next section introduces the general QR and the scale-location mixture representation to cast the model in terms of a standard generalized additive regression with auxiliary latent variables. We then focus on the different priors used, provide additional details on the nonlinear components of the models, briefly discuss VB, outline how we estimate the posterior distributions of the parameters and latent quantities, and illustrate the computational properties of our approach. Section \ref{sec: appl} discusses our empirical findings while Section \ref{extensions} discusses other applications and extensions of the models we develop in this paper. The final section summarizes and concludes the paper. An Online Appendix includes additional technical details, empirical results and more precise information on the used dataset.

\section{Bayesian analysis of general QRs}
\subsection{The likelihood function}
In this paper, our goal is to model the dependence between the $\mathfrak{q}^{th}$ quantile of GDP growth $y_t$ and a  panel of $K$ predictors in $\{\bm x_t\}_{t=1}^T$ with $K$ being huge. The covariates include a wide range of macroeconomic and financial indicators. Possible nonlinearites between $y_t$ and $\bm x_t$ are captured through a function $g_\mathfrak{q}(\bm x_t)$, with $g_\mathfrak{q}: \mathbb{R}^K \to \mathbb{R}$. The fact that $K$ is large and the inclusion of nonlinear functions of $\bm x_t$ implies that the number of parameters is large relative to the number of observations $T$.

Our workhorse model is the QR developed in  \cite{koenker1978regression}. As opposed to the standard QR, our model decomposes the $\mathfrak{q}^{th}$ quantile function $\mathcal{Q}_{\mathfrak{q}(y_t)}$ into a linear and  nonlinear part and a non-standard error distribution:
\begin{equation}
y_t = \bm x'_t \bm \beta_\mathfrak{q} + g_\mathfrak{q}(\bm x_t)+\varepsilon_t,
\end{equation} 
where $\bm \beta_\mathfrak{q}$ is a $K-$dimensional vector of quantile-specific regression coefficients and $\varepsilon_t$ is a shock term with density $f_\mathfrak{q}$ such that the $\mathfrak{q}^{th}$ quantile equals zero:
\begin{equation*}
\int_{-\infty}^0 f_\mathfrak{q}(\varepsilon_t) d \varepsilon_t = \mathfrak{q}.
\end{equation*}
Conditional on the quantile, this model resembles a generalized additive model (GAM), see \cite{hastie1987generalized}. This specification differs from much of the literature \citep[e.g.,][]{adrian2019vulnerable,kohns2021decoupling,mitchell2022nowcasting,carriero2022specification} which sets $g_\mathfrak{q}(\bm x_t) = 0$ for all $t$.

We  approximate $g_\mathfrak{q}(\bm x_t)$ using nonlinear transformations of  $\bm x_t$:
\begin{equation}
g_\mathfrak{q}(\bm x_t) \approx \sum_{m=1}^M \gamma_{\mathfrak{q} m} z_m(\bm x_t) = \bm z'_{t}\bm \gamma_\mathfrak{q} \label{eq:nonlinear}
\end{equation}
with $\bm \gamma_{\mathfrak{q}} = (\gamma_{\mathfrak{q} 1}, \dots, \gamma_{\mathfrak{q} M})'$, $\bm z_t = (z_{1}(\bm x_t), \dots, z_{M}(\bm x_t))'$ and  
$z_{m}(\bm x_t)$ denotes a basis function that depends on $\bm x_t$ with $\gamma_{\mathfrak{q} m}$ denoting the corresponding basis coefficient. This basis function depends on the specific approximation model used to infer the nonlinear effects and our additive representation nests models commonly used in the machine learning literature (such as Gaussian processes, splines, neural networks but also more traditional specifications such as time-varying parameter models). We will discuss the precise specification of $z_m$ (and thus $\bm z_t$) in more detail in Subsection \ref{sec: nonlinear}. Here it suffices to note that depending on the specification, $M$ could be very large. For instance, in the Gaussian process case, $M=T$ and thus the number of regression coefficients would be $K + T$.

If $f_\mathfrak{q}$ remains unspecified, estimation of $\bm \beta_\mathfrak{q}$ and $\bm \gamma_\mathfrak{q}$ is achieved by solving the following optimization problem:
\begin{equation*}
\argmin_{\{\bm \beta_\mathfrak{q}, \bm \gamma_\mathfrak{q}\}} = \sum_{t=1}^{T} \rho_\mathfrak{q} (y_t - \bm x'_t \bm \beta_\mathfrak{q} - \bm z'_t \bm \gamma_\mathfrak{q}),
\end{equation*}
with $\rho_\mathfrak{q}(l) = l [ \mathfrak{q} - \mathbb{I}( l < 0)]$ denoting the loss function.  This optimization problem is straightforward to solve but, if $K+M$ is large, regularization is necessary. This motivates a Bayesian approach to estimation and inference.

From a Bayesian perspective, carrying out posterior inference requires the specification of a likelihood and suitable priors. Following \cite{yu2001bayesian} we assume that the shocks $\varepsilon_t$ follow an asymmetric Laplace distribution (ALD) with density:
\begin{equation*}
f_\mathfrak{q}(\varepsilon_t) = \mathfrak{q} (1-\mathfrak{q}) \exp{(- \rho_\mathfrak{q} (\varepsilon_t))}.
\end{equation*}
The key thing to notice is that the $\mathfrak{q}^{th}$ quantile equals zero and the parameter $\mathfrak{q}$ controls the skewness of the distribution. \cite{kozumi2011gibbs} show that one can introduce auxiliary latent quantities to render the model with ALD distributed shocks conditionally Gaussian. This is achieved by exploiting a scale-location mixture representation \citep{west1987scale}:
\begin{align*}
\varepsilon_{\mathfrak{q} t} &= \theta_\mathfrak{q} \nu_{\mathfrak{q} t} + \tau_\mathfrak{q} \sqrt{\sigma_\mathfrak{q} \nu_{\mathfrak{q} t}} u_t, \\
\theta_\mathfrak{q} &= \frac{1-2\mathfrak{q}}{\mathfrak{q}(1-\mathfrak{q})}, \quad \tau_\mathfrak{q}^2 = \frac{2}{\mathfrak{q}(1-\mathfrak{q})}, \quad \nu_{\mathfrak{q} t} \sim \mathcal{E}\left(\frac{1}{\sigma_\mathfrak{q}}\right), \quad u_t \sim \mathcal{N}(0, 1),
\end{align*}
where $ \mathcal{E}\left(\frac{1}{\sigma_\mathfrak{q}}\right)$ denotes the exponential distribution and $\sigma_\mathfrak{q}$ is a scaling parameter. Hence, conditional on knowing $\nu_\mathfrak{q}=(\nu_{\mathfrak{q} 1}, \dots, \nu_{\mathfrak{q} T})', \theta_\mathfrak{q}, \tau_\mathfrak{q}$, $\sigma_\mathfrak{q}$ and appropriately selecting $g_\mathfrak{q}$, the model is a linear regression model with response $\hat{y_t}= y_t - \theta_\mathfrak{q} \nu_{\mathfrak{q}t}$ and Gaussian shocks that are conditionally heteroskedastic. This conditional likelihood will form the basis of our estimation strategy. 

To complete the model specification we assume that $\frac{1}{\sigma_\mathfrak{q}} \sim \mathcal{G}(c_0,d_0)$, where $c_0$ is the shape and $d_0$ is the rate parameter of the Gamma distribution. Both to zero in order to obtain a flat prior. The choice of the prior distribution on $\bm \beta_\mathfrak{q}$ and $\bm \gamma_\mathfrak{q}$ is essential for our high dimensional QRs. We discuss different suitable choices in the next section.
 
\subsection{Priors for the quantile regression coefficients}
For the large datasets we consider in this paper, $M+K \gg T$ and thus suitable shrinkage priors are necessary to obtain precise inference. \cite{kohns2021decoupling} and \cite{mitchell2022nowcasting} use flexible shrinkage priors in large-scale QRs and show that these work well for tail forecasting. We build on their findings by considering  a range of different priors on $\bm \beta_\mathfrak{q}$ and $\bm \gamma_\mathfrak{q}$. All these priors belong to the class of so-called global-local shrinkage priors \citep{polson2010shrink} and have the following general form:
\begin{align*}
\bm \beta_\mathfrak{q}|\psi^\beta_{\mathfrak{q} 1}, \dots, \psi^\beta_{\mathfrak{q} K}, \lambda^\beta_\mathfrak{q} &\sim \prod_{j=1}^K \mathcal{N}(0, \psi_{\mathfrak{q} j}^{\beta} \lambda^\beta_\mathfrak{q}), \quad \psi^\beta_{\mathfrak{q} j}\sim u, \quad \lambda^\beta_\mathfrak{q} \sim \pi,\\
\bm \gamma_\mathfrak{q}|\psi^\gamma_{\mathfrak{q} 1},\dots , \psi^\gamma_{\mathfrak{q} M}, \lambda^\gamma_\mathfrak{q} &\sim \prod_{j=1}^M \mathcal{N}(0, \psi_{\mathfrak{q} j}^{\gamma} \lambda^\gamma_\mathfrak{q}), \quad \psi^\gamma_{\mathfrak{q} j}\sim u, \quad \lambda^\gamma_\mathfrak{q} \sim \pi,
\end{align*}
with $\lambda_\mathfrak{q}^s~(s \in \{\beta, \gamma\})$ denoting a quantile-specific global shrinkage parameter and $\psi^s_{\mathfrak{q} j}$ are local scaling parameters that allow for non-zero coefficients in the presence of strong global shrinkage (i.e., with $\lambda_\mathfrak{q}^s$ close to zero). The functions $u$ and $\pi$ refer to  mixing densities which, if suitably chosen, translate into different shrinkage priors. In this paper, all the priors we consider can be cast into this form but differ in the way the mixing densities $u$ and $\pi$ are chosen. Since these priors are well known, we briefly discuss them in the main text and relegate additional technical details to the Online Appendix.

We focus on six shrinkage priors that have been shown to work well in a wide variety of forecasting applications \citep[see, e.g.,][]{huber2019adaptive, cross2020macroeconomic, chan2021minnesota, prueser2022}. 

The first prior we consider is the {Ridge prior}. The Ridge prior is a special case of a global-local prior with local parameters set equal to $1$ and a global shrinkage parameter which follows an inverse Gamma distribution. Formally, this implies setting $\psi^s_{\mathfrak{q} j}=1$ for all $\mathfrak{q}, j$ and $\lambda^s_\mathfrak{q} \sim \mathcal{G}^{-1}(e_0, e_1)$. The hyperparameters $e_0$ and $e_1$ control the tightness of the prior. We set these equal to $e_0 = e_1 = 0.01$.  This prior shrinks all coefficients uniformly towards zero and provides little flexibility to allow for idiosyncratic (i.e., variable-specific) deviations from the overall shrinkage pattern.

This issue is solved by estimating the local shrinkage parameters. The Horseshoe \citep[HS, see,][]{carvalho2010horseshoe}, our second prior, does this. This prior sets $u$ and $\pi$ to a half-Cauchy distribution: $\sqrt{\psi^s_{\mathfrak{q} j}} \sim \mathcal{C}^+(0, 1)$ and $\sqrt{\lambda^s_\mathfrak{q}} \sim \mathcal{C}^+(0, 1)$. The HS possesses excellent posterior contraction properties \citep[see, e.g.,][]{horseshoe1,horseshoe2,horseshoe3}. Moreover, it  does not rely on any additional tuning parameters.

Another popular global-local shrinkage prior is the {Normal-Gamma} (NG) prior of \cite{brown2010inference}. This prior assumes that $u$ and $\pi$ are Gamma densities. More formally, $\psi^s_{\mathfrak{q} j} \sim  \mathcal{G}(\vartheta, \lambda^s_\mathfrak{q} \vartheta/2)$ and $\lambda^s_\mathfrak{q}\sim  \mathcal{G}(c_0, d_0)$, with $\vartheta$ being a hyperparameter that controls the tail behavior of the prior, and $c_0$ and $d_0$ are hyperparameters that determine the overall degree of shrinkage. We set $c_0 = d_0 = 0.01$ and $\vartheta=0.1$. This choice implies heavy global shrinkage on the coefficients but also implies fat tails of the marginal prior of the coefficients after integrating out the local scaling parameters. The Bayesian LASSO is obtained as a special case of the NG prior with $\vartheta=1$.

 Next, we consider the Dirichlet-Laplace prior \citep{dirlaplace} that assumes the local scaling parameter $\psi^s_{\mathfrak{q} j}$ is a product of a Dirichlet-distributed random variate $\phi^s_{\mathfrak{q} j} \sim \text{Dir}(\alpha, \dots, \alpha)$ and a parameter $\tilde{\psi^s_{\mathfrak{q} j}} \sim \mathcal{E}(1/2)$ that follows an exponential distribution. Hence, the Dirichlet-Laplace prior sets $\psi^s_{\mathfrak{q} j} = (\phi^s_{\mathfrak{q} j})^2 \tilde{\psi^s_{\mathfrak{q} j}}$.  On the global scaling parameters we use a Gamma distribution $\sqrt{\lambda^\beta_\mathfrak{q}} \sim \mathcal{G}(K \alpha, 1/2)$ and $\sqrt{\lambda^\gamma_\mathfrak{q}} \sim \mathcal{G}(M \alpha, 1/2)$. We set $\alpha =\frac{1}{K}$ for the linear term and $\alpha =\frac{1}{M}$ for the non-linear term.

 Finally, we also consider the Minnesota prior as implemented in \cite{carriero2022specification}. This prior assumes that the local scaling parameters are $\psi^s_{\mathfrak{q} j}= \frac{\sigma^2_y}{\sigma_i^2}$, where $\sigma^2_y$ is the OLS variance of a standard linear regression model, and $\sigma_i^2$ is the error variance of a linear regression with the $i^{th}$ variable used as the dependent variable. This term serves to control for scaling differences. The global parameter $\lambda^s_\mathfrak{q} = \lambda_1 \lambda_2$ is often fixed and depends on two terms. The first term ($\lambda_1$) controls the overall shrinkage of the prior, whereas the second term ($\lambda_2$) controls the rate of shrinkage on variables other than lagged GDP growth. $\lambda_2$ is then often set to values smaller than $1$ if $i$ marks a variable other than lagged GDP growth or it equals $1$ if it is related to the first lag of GDP growth. \cite{carriero2022specification} suggest using $\lambda_1 = 0.04$ and $\lambda_2 = 0.25$ and find that this choice leads to accurate tail forecasts. However, given that our models are substantially larger, we also include a version of the prior that estimates $\lambda_1$ and $\lambda_2$ by placing Gamma priors on both of them. We will refer to it as the Minnesota-Gamma (Minn-Gamma) prior.\footnote{It is worth stressing that this hierarchical version of the Minnesota prior is closely related to the Ridge prior above if the covariates in $\bm x_t$ are standardized to have unit variance.}

\subsection{Capturing nonlinearities in high dimensional QRs}\label{sec: nonlinear}
In extreme periods such as the GFC or the Covid-19 pandemic, nonlinearities in macroeconomic data become prevalent.  We control for this by having a nonlinear part in our QR. As stated in eq. (\ref{eq:nonlinear}), we capture possible nonlinearities in $\bm x_t$ through nonlinear transformations $z_m(\bm x_t)$. 

The first and simplest nonlinear specification  maps $\bm x_t$ into the space of polynomials. \cite{bai2008forecasting} capture nonlinearities in macro data through polynomials relying on factor-based predictive regressions. We follow this approach and define the corresponding basis function as follows: 
\begin{equation*}
    \bm z_t = ((\bm x^2_t)', (\bm x^3_t)', \dots, (\bm x^N_t)')'.
\end{equation*}
Deciding on the order of the polynomial $N$ is a model selection issue and suitable shrinkage priors can be adopted. In our empirical work, we focus on the cubic case. This specification will overweight large movements in $\bm x_t$ and should thus be suitable for quickly capturing sharp downturns in the business cycle. In this case, the number of basis coefficients is $M=3K$. The resulting nonlinear model is called \textit{Polynomial-QR}.


Adding cubic terms allows us to capture nonlinearities in a relatively restricted manner. Since the precise form of nonlinearities is typically unknown, the remaining two specifications we consider are nonparametric and only require relatively mild prior assumptions on the form of nonlinear interactions. The first of these two is the B-Spline \citep[see, e.g.,][for a review]{DeBoor2001}. B-Splines have a proven track record in machine learning and computer science \citep{shin2020functional}. 

For the B-spline,  we assume that each element in $\bm x_t$ exerts a (possibly) nonlinear effect on $y_{t}$ that might differ across covariates. This implies that $g_\mathfrak{q}(\bm x_t)$ equals:
\begin{equation*}
    g_\mathfrak{q}(\bm x_t) \approx \sum_{k=1}^K \bm \Phi_k(\bm x_{\bullet, j}) \bm \gamma_{\mathfrak{q}, k}.
\end{equation*}
Here, we let $\bm \Phi_k$ denote a $T \times r$ matrix of B-spline basis functions that depend on the $j^{th}$ covariate in $\bm X = (\bm x'_1, \dots, \bm x'_T)'$, $\bm x_{\bullet, j}$, and $r$ is the number of knots. In this case, the number of nonlinear coefficients is $M=rK$. In our empirical work, we place the knots at the following quantiles of $\bm x_{\bullet, j}$: \{0, 0.05, 0.1, 0.25, 0.50, 0.75, 0.90, 0.95, 1\}, implying that $r=9$ and thus $M=9K$. We will henceforth call this model \textit{Spline-QR}.

The last specification we consider is the Gaussian process (GP) regression. GP regression is a nonparametric estimation method that places a GP prior on the function  $g_\mathfrak{q}(\bm x_t)$:
\begin{equation*}
    g_\mathfrak{q}(\bm x_t) \sim \mathcal{GP}(\mu_\mathfrak{q}(\bm x_t), \mathcal{K}(\bm x_t, \bm x_\mathfrak{t} )).
\end{equation*}
The mean function $\mu_\mathfrak{q}(\bm x_t)$ is, without loss of generality, set equal to zero and $\mathcal{K}(\bm x_t, \bm x_\mathfrak{t})$ is a kernel function that encodes the relationship between $\bm x_t$ and $\bm x_\mathfrak{t}$ for $t, \mathfrak{t} = 1, \dots, T$. It is worth noting that our additive specification implies that if the mean function is set equal to zero, the model is centered on a standard QR. 

Since $\bm x_t$ is observed in discrete time steps, the GP prior implies a Gaussian prior on $\bm g_\mathfrak{q} = (g_\mathfrak{q}(\bm x_1), \dots, g_\mathfrak{q}(\bm x_T))'$:
\begin{equation*}
    \bm g_\mathfrak{q} \sim \mathcal{N}(\bm 0_T, \bm K(\bm w)),
\end{equation*}
where $\bm K(\bm w)$ is a $T \times T$-dimensional matrix with $(t, \mathfrak{t})^{th}$ element $\mathcal{K}(\bm x_t, \bm x_\mathfrak{t})$. $\bm w =(w_1, w_2)'$ is a set of hyperparameters that determine the properties of the kernel (and thus the estimated function). 

The GP regression is fully specified if we determine the kernel function $\mathcal{K}$. In this paper, we use the Gaussian (or squared exponential) kernel:
\begin{equation*}
    \mathcal{K}(\bm x_t, \bm x_\mathfrak{t}) = w_1 \times \exp\left(- \frac{w_2}{2}||\bm x_t - \bm x_\mathfrak{t}||^2 \right).
\end{equation*}
The hyperparameters $\bm w$ are set according to the median heuristic proposed in \cite{Arin2017}. 

What we discuss above is the function-space view of the GP regression. An alternative way of expressing the GP is the so-called weight-space view. The weight-space view is obtained by integrating out $\bm g_\mathfrak{q}$, yielding the following regression representation:
\begin{equation*}
    \bm y = \bm X \bm \beta_\mathfrak{q}  + \bm Z \bm \gamma_\mathfrak{q} + \bm \varepsilon,
\end{equation*}
with $\bm y$ denoting the stacked dependent variables, $\bm Z$ is the lower Cholesky factor of $\bm K$ and $\bm \gamma_\mathfrak{q} \sim \mathcal{N}(0, \bm I_T)$. Notice that $\bm g_\mathfrak{q} = \bm Z \bm \gamma_\mathfrak{q}$. Hence, the Cholesky factor of the kernel matrix provides the basis functions, and the parameters can be readily estimated. In this case, the number of nonlinear coefficients is $M=T$. Since we use a shrinkage prior on $\bm \gamma_\mathfrak{q}$, the corresponding implied kernel is given by $\bm Z \bm B^\gamma_\mathfrak{q} \bm Z'$. The $M \times M$ matrix $\bm B^\gamma_\mathfrak{q}$ is a prior covariance matrix with $\bm B^\gamma_\mathfrak{q} = \lambda_\mathfrak{q}^\gamma \times \text{diag}(\psi^\gamma_{\mathfrak{q}1}, \dots, \psi^\gamma_{\mathfrak{q}M})$.  Approximating $g_\mathfrak{q}$ using GPs leads to the \textit{GP-QR} specification.

This completes our choice of nonlinear techniques used in the big data QR. Alternative choices (such as allowing for time-varying parameters, neural networks or Bayesian additive regression trees) can be straightforwardly introduced in this general framework.

\subsection{A brief introduction to variational Bayes}
The high dimensionality of the state space calls for alternative techniques to carry out posterior inference. We opt for using variational approximations to the joint posterior density. In this section, we provide a discussion on how VB works in general. For an excellent in-depth introduction, see \cite{blei2017variational}. In machine learning, variational techniques have been commonly used to estimate complex models such as deep neural networks \citep[see, e.g.,][]{polson2017deep}. In econometrics, recent papers use VB in huge dimensional multivariate time series models such as VARs  \citep{gefang2022forecasting, chan2020fast} or state space models to speed up estimation \citep{koop2018variational}. In a recent paper, \cite{korobilis2022probabilistic}, propose a QR factor model and estimate it using VB techniques.

To simplify the exposition, we fix the prior variances. The appendix provides information on how we estimate the prior variances (and associated hyperparameters) using VB. Let $\bm \xi_\mathfrak{q} = (\bm \beta_\mathfrak{q}, \bm \gamma_\mathfrak{q}, \sigma_\mathfrak{q}, \bm \nu_\mathfrak{q})$ denote a generic vector which stores all unknowns of the model, with $\bm \nu_\mathfrak{q} = (\nu_{\mathfrak{q} 1}, \dots, \nu_{\mathfrak{q} T})$ denoting the latent components. 

Our aim is to approximate the joint posterior distribution $p(\bm \xi_\mathfrak{q}|\bm{y})$ using an analytically tractable approximating distribution $q(\bm \xi_\mathfrak{q})$. This variational approximation is found by minimizing the Kullback-Leibler (KL) distance between $p$ and $q$. One can show that minimization of the KL distance is equivalent to maximizing the evidence lower bound (ELBO) defined as:
\begin{equation}
\text{ELBO} = \mathbb{E}_{q(\bm \xi_\mathfrak{q})} \left( \log p(\bm \xi_\mathfrak{q}, \bm y)\right) - \mathbb{E}_{q(\bm \xi_\mathfrak{q})} \left(\log q(\bm \xi_\mathfrak{q})\right),
\end{equation}
with $\mathbb{E}_{q(\bm \xi_\mathfrak{q})}$ denoting the expectation with respect to $q(\bm \xi_\mathfrak{q})$. This implies that finding the approximating density $q$ replaces the integration problem (which is typically solved through MCMC sampling) with an optimization problem (which is fast and thus scales well into high dimensions). 

A common and analytically tractable choice of approximating densities assumes that $q(\bm \xi_\mathfrak{q})$ is factorized as follows:
\begin{equation*}
q(\bm \xi_\mathfrak{q}) = \prod_{s=1}^S q_s(\bm \xi_{\mathfrak{q} s}),
\end{equation*}
where $\bm \xi_{\mathfrak{q} s}$ denotes a partition of $\bm \xi_\mathfrak{q}$. A particular example (which we use in this paper) would specify $\bm \xi_{\mathfrak{q} 1} = (\bm \beta'_\mathfrak{q}, \bm \gamma'_\mathfrak{q})',$ $\xi_{\mathfrak{q} 2} = \sigma_\mathfrak{q}$ and $\bm \xi_{\mathfrak{q} 3}  = \bm \nu_\mathfrak{q}$. 

This class is called the mean field variational approximation and assumes that the different blocks $\bm \xi_{\mathfrak{q} s}$ are uncorrelated.\footnote{\cite{Frazier2022} state that mean field VB approximations might perform poorly in models with a large number of latent variables. However, they also note that the resulting model forecasts could still perform well in practice.} Notice that all our priors on $\bm \xi_\mathfrak{q}$ can be written as:
\begin{equation*}
p(\bm \xi_\mathfrak{q}) = \prod_{s=1}^S p(\bm \xi_{\mathfrak{q} s}),
\end{equation*}
and using the fact that:
\begin{equation*}
\mathbb{E}_{q(\bm \xi_\mathfrak{q})}(\log p(\bm \xi_\mathfrak{q}, \bm y)) = \mathbb{E}_{q(\bm \xi_\mathfrak{q})} (\log  p(\bm y | \bm \xi_\mathfrak{q})) + \sum_{s=1}^S \mathbb{E}_{q(\bm \xi_\mathfrak{q})} (\log p(\bm \xi_{\mathfrak{q} s})),
\end{equation*}
the ELBO can be stated as:
\begin{equation*}
\text{ELBO} = \mathbb{E}_{q(\bm \xi_\mathfrak{q})} (\log p(\bm y | \bm \xi_\mathfrak{q})) + \sum_{s=1}^S \mathbb{E}_{q(\bm \xi_\mathfrak{q})} (\log p(\bm \xi_{\mathfrak{q} s})) -  \sum_{s=1}^S \mathbb{E}_{q(\bm \xi_\mathfrak{q})} (\log q(\bm \xi_{\mathfrak{q} s})).
\end{equation*}
\cite{wand2011mean} prove that under the variational family the optimal approximating densities are closely related to the full conditional posterior distributions:
\begin{equation*}
q^*_s(\bm \xi_\mathfrak{q}) = \exp \left[\mathbb{E}_{q(\bm \xi_\mathfrak{q})} ( \log p(\bm \xi_{\mathfrak{q} s} | \bm y, \bm \xi_{\mathfrak{q}, -s})\right],
\end{equation*}
where $\bm \xi_{\mathfrak{q}, -s}$ is the vector $\bm \xi_\mathfrak{q}$ with the $s^{th}$ component excluded. Hence, if $p(\bm \xi_{\mathfrak{q} s} | \bm y, \bm \xi_{\mathfrak{q}, -s})$ is known (which is the case for the QR regression based on the auxiliary representation discussed in the previous subsection), the elements in $\bm \xi_{\mathfrak{q} s}$ can be updated iteratively (by conditioning on the expected values of $\bm \xi_{\mathfrak{q}, -s}$) until the squared difference of the ELBO or of all elements of $\xi_{\mathfrak{q} s}$ is smaller than some small $\epsilon$ between two subsequent iterations.

\subsection{Approximate Bayesian inference in general QRs}
In this section, we briefly state the three approximating densities $(q^*_s(\bm \xi))$ used to estimate the parameters and latent quantities in the QR regression.  We provide derivations for the three approximating densities of the three parameter groups: $\tilde{\bm \beta}_\mathfrak{q} = (\bm \beta'_\mathfrak{q}, \bm \gamma'_\mathfrak{q})',$ $\sigma_\mathfrak{q}$ and $\bm \nu_\mathfrak{q}$ in the Online Appendix. 

We start by discussing the approximating densities for the regression and basis coefficients. A Gaussian distribution approximates the posterior of $\tilde{\bm \beta}_\mathfrak{q}$:
\begin{equation*}
    p(\tilde{\bm \beta}_\mathfrak{q} |\bullet) \approx \mathcal{N}\left(\mathbb{E}(\tilde{\bm \beta}_\mathfrak{q}), \hat{\bm{\Sigma}}_{\bm{\kappa}_\mathfrak{q}}\right),
\end{equation*}
with variance and mean given by, respectively:
\begin{align*}
\hat{\bm{\Sigma}}_{\tilde{\bm \beta}_\mathfrak{q} } &= \left[ \sum_{t=1}^T \frac{\bm{f}_t \bm{f}_t'}{\tau_\mathfrak{q}^2} \mathbb{E}\left(\frac{1}{\nu_{\mathfrak{q}t}}\right)\mathbb{E}\left(\frac{1}{\sigma_\mathfrak{q}}\right) + \bm{B}_{0 \mathfrak{q}}^{-1}\right]^{-1},  \\
\mathbb{E}(\tilde{\bm \beta}_\mathfrak{q}) &=\hat{\bm{\Sigma}}_{\tilde{\bm \beta}_\mathfrak{q} } \left[ \mathbb{E}\left(\frac{1}{\sigma_\mathfrak{q}}\right) \sum_{t=1}^T \mathbb{E}\left(\frac{1}{\nu_{\mathfrak{q} t}}\right) \frac{\bm{f}_t\left(y_t - \theta_\mathfrak{q} \left[\mathbb{E}\left(\frac{1}{\nu_{\mathfrak{q} t}}\right)\right]^{-1}\right)}{\tau_\mathfrak{q}^2}\right].\label{normalsigma_1}
\end{align*}
$\bm f_t = (\bm x'_t, \bm z'_t)'$ and $\bm{B}_{0 \mathfrak{q}}^{-1} =  \text{diag}(\bm B^\beta_\mathfrak{q}, \bm B^\gamma_\mathfrak{q})^{-1}$ is a prior precision matrix with $\bm B^\beta_\mathfrak{q} =\lambda^\beta_\mathfrak{q}  \times  \text{diag}(\psi^\beta_{\mathfrak{q} 1}, \dots, \psi^\beta_{\mathfrak{q} K})$ and $\bm B^\gamma_\mathfrak{q} =\lambda^\gamma_\mathfrak{q}  \times  \text{diag}(\psi^\gamma_{\mathfrak{q} 1}, \dots, \psi^\gamma_{\mathfrak{q} K})$. The approximating densities used to estimate the prior hyperparameters are provided in Section 1 of the Online Appendix. 

The latent variable $\nu_{\mathfrak{q} t}$ follows a generalized inverse Gaussian (GIG) distribution: $\text{GIG}(r, A, B)$\footnote{We use the following parametrization of the GIG distribution: $\log\left(\text{GIG(x)}\right) \propto (0.5-1)\log(x) - \left(Ax + \frac{1}{2} \frac{B}{x}\right)  $.}  with 

\begin{equation*}
    p(\nu_{\mathfrak{q} t}|\bullet) \approx \text{GIG}\left(\frac{1}{2},~ \underbrace{2\mathbb{E} \left(\frac{1}{\sigma_\mathfrak{q}}\right) + \frac{\theta_\mathfrak{q}^2}{\tau_\mathfrak{q}^2} \mathbb{E}\left(\frac{1}{\sigma_\mathfrak{q}}\right)}_{A_\mathfrak{q}},\quad \underbrace{\frac{\mathbb{E}\left(\frac{1}{\sigma_\mathfrak{q}}\right)}{\tau_\mathfrak{q}^2} \left[ \left(y_t - \bm{f}_t'\mathbb{E}(\tilde{\bm{\beta}}_\mathfrak{q})\right)^2 + \bm{f}_t' \hat{\bm \Sigma}_{\tilde{\bm{\beta}}_\mathfrak{q}}\bm{f}_t\right]}_{B_\mathfrak{q}}\right).
\end{equation*}
The moments of $\nu_{\mathfrak{q} t}$ are given by
\begin{equation*}
\mathbb{E}\left(\nu_{\mathfrak{q} t}^j\right) = \left(\frac{\sqrt{B_\mathfrak{q}}}{\sqrt{A_\mathfrak{q}}}\right)^j \frac{K_{1/2+j}\left(\sqrt{A_\mathfrak{q}B_\mathfrak{q}}\right)}{K_{1/2}\left(\sqrt{A_\mathfrak{q} B_\mathfrak{q}}\right)},
\end{equation*}
where $K_{x}$ denotes the modified Bessel function of the second kind. 

Finally, we approximate 
\begin{equation*}
   p\left(\frac{1}{\sigma_\mathfrak{q}}|\bullet\right) \approx \mathcal{G}(c_{\mathfrak{q} 1}, d_{\mathfrak{q} 1}) 
\end{equation*}
with 
\begin{align*}
c_{\mathfrak{q} 1} &= c_0 + 1.5T ,\\
d_{\mathfrak{q} 1} &=d_0 + \sum_{t=1}^T \mathbb{E}(\nu_{\mathfrak{q} t}) + \frac{1}{2\tau_\mathfrak{q}^2} \sum_{t=1}^T (\mathbb{E}\left(\frac{1}{\nu_{\mathfrak{q} t}}\right)\left(y_t - \bm{f}_t'\mathbb{E}(\tilde{\bm{\beta}}_\mathfrak{q})\right)^2 + 2 \theta_\mathfrak{q} (\bm{f}_t'\mathbb{E}(\tilde{\bm{\beta}}_\mathfrak{q})-y_t) \nonumber \\
&+\mathbb{E}(\nu_{\mathfrak{q} t})\theta_\mathfrak{q}^2 + \mathbb{E}\left(\frac{1}{\nu_{\mathfrak{q} t}}\right)\bm{f}_t' \hat{\bm{\Sigma}}_{\tilde{\bm{\beta}}_\mathfrak{q}} 
 \bm{f}_t), \label{eqbeta}
\end{align*}
and $\mathbb{E}\left(\frac{1}{\sigma_\mathfrak{q}}\right) = \frac{c_{\mathfrak{q} 1}}{d_{\mathfrak{q} 1}}$.

\subsection{Comparing computation times between VB and MCMC}
The steps for updating $\bm \xi_\mathfrak{q}$, in combination with the updating steps for the priors detailed in the Online Appendix, form the basis of our VB algorithm. As stated in the introduction, the key advantage of using VB instead of more precise MCMC-based techniques is computational efficiency. Before we turn to our empirical work, we illustrate this point using synthetic data.

\begin{figure}[h!]
\centering 
\includegraphics[trim = 0mm 0mm 0mm 8mm, clip,scale=0.6]{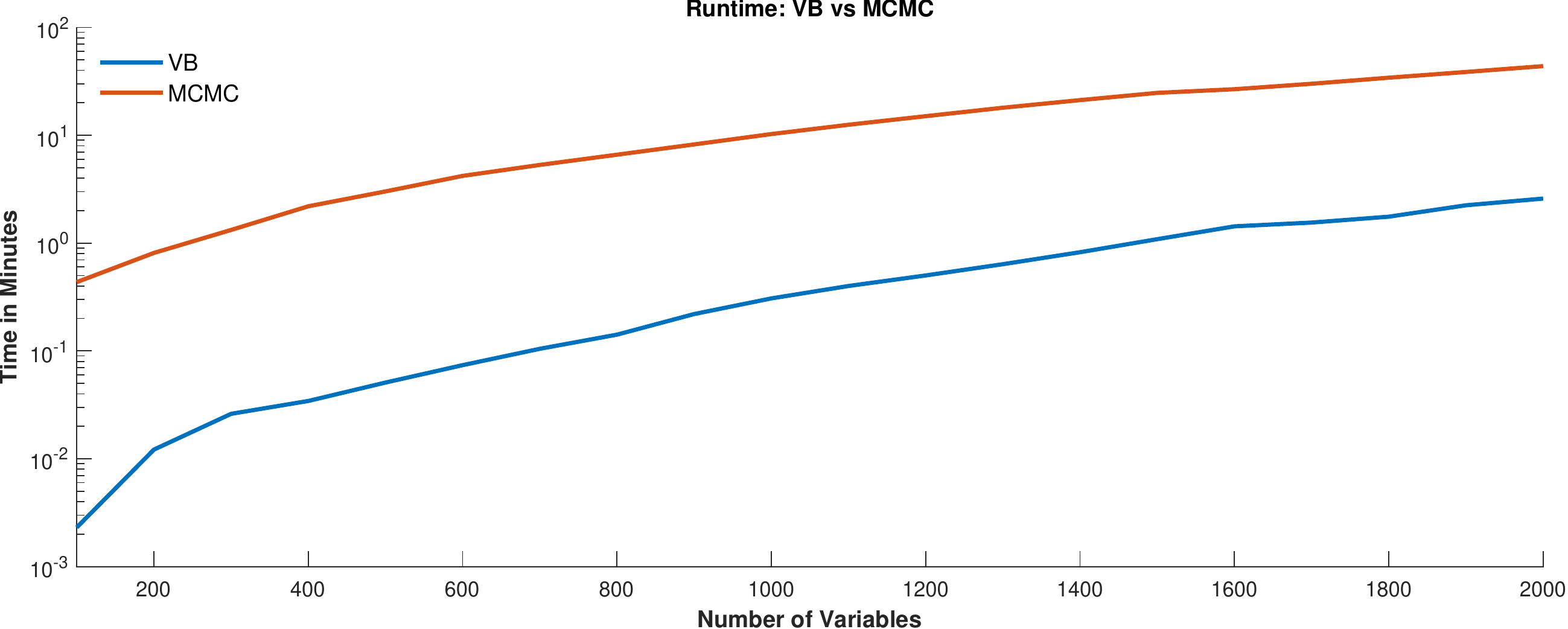} 
\caption[]{Comparison of computation times against the number of covariates $M+K$. Computation times are on the logarithmic scale.}
\label{fig:Runetime_revision} 
\end{figure}

To illustrate the computational merits of employing VB-based approximations, Fig. \ref{fig:Runetime_revision} shows the estimation times (in logarithmic scale) for different values of $M+K$ using our VB-based QR (for a specific quantile) and the QR estimated through the Gibbs sampler. The MCMC algorithm is repeated $10,000$ times.
The figure shows that the computational burden increases lightly in the number of covariates for VB. When we focus on MCMC estimation, the computational requirements increase sharply in the number of covariates. Especially in our empirical work, where $K+M$ is often above $1,000$, VB proves to be a fast alternative to MCMC-based quantile regressions.  MCMC-based estimation (without leveraging parallel computing facilities) becomes excessively slow (or even infeasible) for forecasting applications (where models have to be estimated sequentially across forecast origins) and if the researcher wishes to estimate a large number of quantiles. In our case, all computations in our empirical work can be carried out on a standard desktop computer. This implies that the models we focus on in this paper can be used for producing forecasts in a timely manner.

\section{Forecasting output growth using huge dimensional QRs}\label{sec: appl}
In this section, we present our forecasting results. The next  subsection provides information on the dataset and the forecasting setup. We then proceed by discussing the results from  QRs that exclude the nonlinear part in Sub-section \ref{sec: linear}. The question whether nonlinearities are important is investigated in Subsection \ref{sec: nonlinearities}, and Sub-section  \ref{ssub:heterogeneity_of_forecast_accuracy_over_time} deals with how forecast accuracy changes over time. Subsection \ref{sec: properties_fcst} discusses the determinants of the different tail forecasts and differences in the shrinkage properties across priors.

\subsection{Data overview and forecasting setup}\label{sec: data}
We use the quarterly version of the \cite{mccrackenngdata} dataset. 
The dataset covers information about the real economy (output, labor, consumption,
orders and inventories), money, prices and financial markets (interest rates, exchange rates, stock market indexes). All series are transformed to be approximately stationary. The set of variables included in $\bm x_t$ and their transformation codes are described in Table 1 of the Online Appendix. All models we consider also include the first lag of GDP growth.\footnote{We find that including more lags of GDP growth only has small effects on the empirical results.} Forecasts are carried out using direct forecasting by appropriately lagging the elements in $\bm x_t$.

Our sample runs from $1971$Q1 to $2021$Q3 and we use  the period  $1991$Q2 to $2021$Q3 as our hold-out period. The forecasting design is recursive. This implies that we estimate all our models on an initial training sample with data until $1991$Q1 and produce one-quarter- and four-quarters-ahead predictive distributions for $1991$Q2 and $1992$Q1, respectively. After obtaining these, we add the next observation ($1991$Q2) and recompute the models to obtain the corresponding predictive densities for $1991$Q3 and $1992$Q2. This procedure is repeated until we reach the end of the hold-out period.

As a measure of overall forecasting accuracy we focus on the continuous ranked probability score (CRPS). The CRPS is a measure of density forecasting accuracy and generalizes the mean absolute error (MAE) to take into account how well a given model predicts higher order moments of a target variable. 

The CRPS measures overall density fit. Considering overall CRPSs possibly masks relevant idiosyncrasies of model performance across quantiles. If a decision maker is interested in downside risks to GDP growth, she might value a model more that does well at the critical $5$ or $10$ percentiles as opposed to the remaining regions of the predictive distribution. To shed light on asymmetries across different predictive quantiles, we focus on the quantile score (QS):
\begin{equation*}
\mbox{QS}_{\mathfrak{q} t} = (y_{t} - Q_{\mathfrak{q} t})(\mathfrak{q} - \mathbf{1}_{\{y_{t} \le
Q_{\mathfrak{q}t}\}}),
\end{equation*}
where $Q_{\mathfrak{q} t}$ is the forecast of the $\mathfrak{q}^{th}$ quantile of $y_t$ and $\mathbf{1}_{\{y_{t} \le
Q_{\mathfrak{q} t}\}}$ denotes the indicator function that equals one if $y_t$ is below the forecast for the $\mathfrak{q}^{th}$ quantile.

The QS can also be used to construct quantile-weighted (qw) CRPS scores \citep{gneiting2011comparing}. These qw-CRPSs can be specified to put more weight on certain regions of the predictive distribution. In general, the qw-CRPS is computed as:
\begin{equation*}
    \text{qw-CRPS} = \frac{2}{J-1} \sum_{j=1}^{J-1} \omega(\zeta_j) \mbox{QS}_{\mathfrak{s}_j t},
\end{equation*}
with $\zeta_j = j/J$, $J-1 = 19$ denoting the number of quantiles we use to set up the qw-CRPS and $\mathfrak{s}_j$ selects the $j^{th}$ element from the set of quantiles we consider. This set ranges from $0.05$ to $0.95$ with a step size of $0.05$ and thus, $\mathfrak{s}_1 = 0.05, \mathfrak{s}_2 = 0.10, \dots, \mathfrak{s}_{19} = 0.95$. 

We use two weighting functions $\omega(\zeta_j)$ that focus on different regions of the predictive density. These schemes are motivated in \cite{gneiting2011comparing}. The first (CRPS-l) puts more weight on the left tail (i.e. downside risks) and is specified as $\omega(\zeta_j) = (1 - \zeta_j)^2$, while the second (CRPS-t) puts more weight on both tails as opposed to the center of the distribution: $\omega(\zeta_j) = (2\zeta_j - 1)^2$. Notice that if we use equal weights, we obtain a discrete approximation to the CRPS.

\subsection{Results based on linear QRs}\label{sec: linear}
We start discussing the QRs that set $g(\bm x_t) = 0$ for all $t$. Since we advocate using VB-based estimation techniques our goal is to show, within feasible dimensions, that our approximation-based approach produces forecast densities of output growth which are competitive to the ones obtained by estimating QRs using MCMC. As a benchmark model we rely on the QR with a Minnesota prior with fixed hyperparameters estimated using MCMC techniques. We use this as a benchmark since \cite{carriero2022specification} show that it displays an excellent overall performance in terms of tail forecasting and is simple to implement.

Then, in addition, we also assess whether the additional information in our dataset translates into predictive gains. This is achieved by also including the model proposed in  \cite{adrian2019vulnerable}. To enable comparison to the original model, we estimate it in the same way as in the original paper and call the model henceforth ABG.


Table \ref{CRPSlinear} shows  average (over time) qw-CRPSs relative to the QR-Minnesota model. Numbers smaller than one suggest that a given model outperforms the  benchmark, whereas numbers exceeding unity indicate that the model produces less precise density forecasts. The upper part of the table shows the results for VB-based approximations, while the lower part of the table displays the results for all models estimated through MCMC sampling.

\begin{table}[t!]
\caption{CRPS for linear models}
\label{CRPSlinear}
\begin{tabular}{lcccc c cccc} 
\toprule
&\multicolumn{4}{ c }{One-quarter-ahead}&&\multicolumn{4}{ c }{Four-quarters-ahead} \\
\cline{2-5} \cline{7-10}
Model   &CRPS&CRPS-t&CRPS-l&CRPS-r&&CRPS&CRPS-t&CRPS-l&CRPS-r\\ \toprule
\multicolumn{4}{l }{VB}& \multicolumn{1}{l }{\textit{   }}&& \\ \midrule
HS& $1.06$ & $1.11$ & $1.12$& $1.03$ && $1.06$ & $1.02$ & $1.07$&$1.02$\\
Ridge&  \A{}$0.88$ & \A{}$0.88$ &\A{}$0.90$&\A{}$0.86$  &&\A{}$0.90$ &\A{} $0.89$ &\B{} $0.90$&\A{}$0.90$\\
NG& $1.01$ & $1.04$ & $1.04$& $1.02$ &&$1.07$ & $1.02$ & $1.08$&1.03\\
LASSO&  \A{}$0.92$ & \B{}$0.93$&$0.98$  &\A{}  $0.86$&&\A{}$0.91$ & \A{}$0.89$ & \B{}$0.91$&\A{}$0.90$\\
DL& $1.12$ & $1.18$ & $1.16$& $1.11$ && $1.16$ & $1.13$ & $1.21$&$1.09$\\
Minn-Gamma& \A{}$0.88$ & \A{}$0.88$ &\A{}$0.90$&\A{}$0.86$   && \A{}$0.90$ &\A{} $0.89$ &\B{} $0.90$&\A{}$0.90$\\ \midrule 
\multicolumn{4}{l }{MCMC}& \multicolumn{1}{l }{\textit{   }}&& \\ \midrule
HS& $1.01$ & $1.03$ & $1.02$& $1.02$ && $1.05$ & $1.06$ & $1.04$&$1.06$\\
Ridge&  \A{}$0.89$ & \A{}$0.89$ &\A{}$0.91$&\A{}$0.86$  &&\A{}$0.87$ &\A{} $0.86$ &\B{} $0.87$&\A{}$0.87$\\
NG& $1.03$ & $1.04$ & $1.01$& $1.06$ &&$1.01$ & $0.96$ & $1.03$&1.00\\
LASSO&  \A{}$0.93$ & \B{}$0.94$&$0.95$  &\A{}  $0.91$&&\B{}$0.97$ & $0.97$ & $0.99$&\A{}$0.94$\\
DL& $1.17$ & $1.15$ & $1.06$& $1.26$ && $1.12$ & $1.11$ & $1.22$&$1.03$\\ 
Minn-Gamma&\A{}$0.89$ & \A{}$0.89$ &\A{}$0.91$&\A{}$0.86$  && \A{}$0.93$ &\A{} $0.92$ & $0.95$&\A{}$0.89$\\  \midrule 
ABG& $1.00$ & $1.04$ & $1.07$& $0.95$ && $1.03$ & $1.03$ & $1.03$&$1.03$\\  \midrule 
\end{tabular} 
\begin{tabular}{p{16cm}}
\footnotesize  \textit{Notes}: We highlight in light gray (dark gray) rejection of equal forecasting accuracy against the benchmark model at significance level 10\% (5\%) using the test in \cite{DieboldMariano1995} with adjustments proposed by \cite{Harvey1997}. Results are shown relative to the Minnesota prior with fixed hyperparameter and are based on the full sample.
\end{tabular}
 \end{table}


We first address the  question of whether VB-based approximations result in less precise forecasts compared to MCMC-based models. The answer to this question is a clear "not at all." The results indicate that the differences in predictive performance between MCMC and VB-based models are minimal, sometimes even approaching zero. While there are instances where VB-based estimates perform slightly worse, such as for the HS, there are also cases where VB-based models surprisingly yield slightly more accurate tail forecasts. Overall, when comparing the upper and lower parts of the table, we find no significant disparities in forecast performance when employing approximations. Furthermore, conducting our extensive forecasting exercise using VB-based models is considerably faster than obtaining predictive densities through MCMC estimation. Even the occasional small losses incurred are negligible considering the ability to generate accurate forecasts of GDP growth rapidly.

When we zoom into the performance of the VB-based models we find a great deal of heterogeneity with respect to different priors. Popular GL priors such as the HS, the NG or the DL lead to forecasts that are often slightly worse than the ones obtained from the benchmark Minnesota QR. However, priors such as the Ridge or the LASSO \citep[which is particularly known for over-shrinking significant signals, see, e.g.,][]{brown2010inference} yield forecasts that are better than the benchmark forecasts for both  horizons and across the different variants of the CRPS. The Minn-Gamma prior yields forecasts which are almost identical to the one of the Ridge. This is not surprising given the fact that the prior introduces strong shrinkage in large dimensions and most coefficients are forced to zero by estimating $\lambda_1$ and $\lambda_2$ to be close to zero. It is also worth stressing that the Minnesota prior with fixed hyperparameters introduces too little shrinkage and leads to overfitting. This points towards the necessity to either specify the global shrinkage parameters to depend on $K$ or estimate them from the data. Finally, when we consider the simple ABG model we  
 corroborate recent results in \cite{carriero2022specification}, who show that large QRs with shrinkage improve upon the ABG model (with improvements depending on the prior). This is especially pronounced in the case of the Ridge prior. In this case, the accuracy gains vis-\'{a}-vis the ABG model reach around 17 percent and,  in most cases, accuracy differences are statistically significant according to the \cite{DieboldMariano1995} test.

Turning to the different forecast horizons reveals that specifications that do well in terms of short-term forecasting also produce precise  longer-term predictions. For the LASSO-based model, four-quarters-ahead accuracy gains are slightly more pronounced whereas for the Ridge we do not find discernible differences across both forecast horizons. 

\begin{figure}[t!]
\centering 
\includegraphics[trim = 0mm 0mm 0mm 8mm,width=1\textwidth]{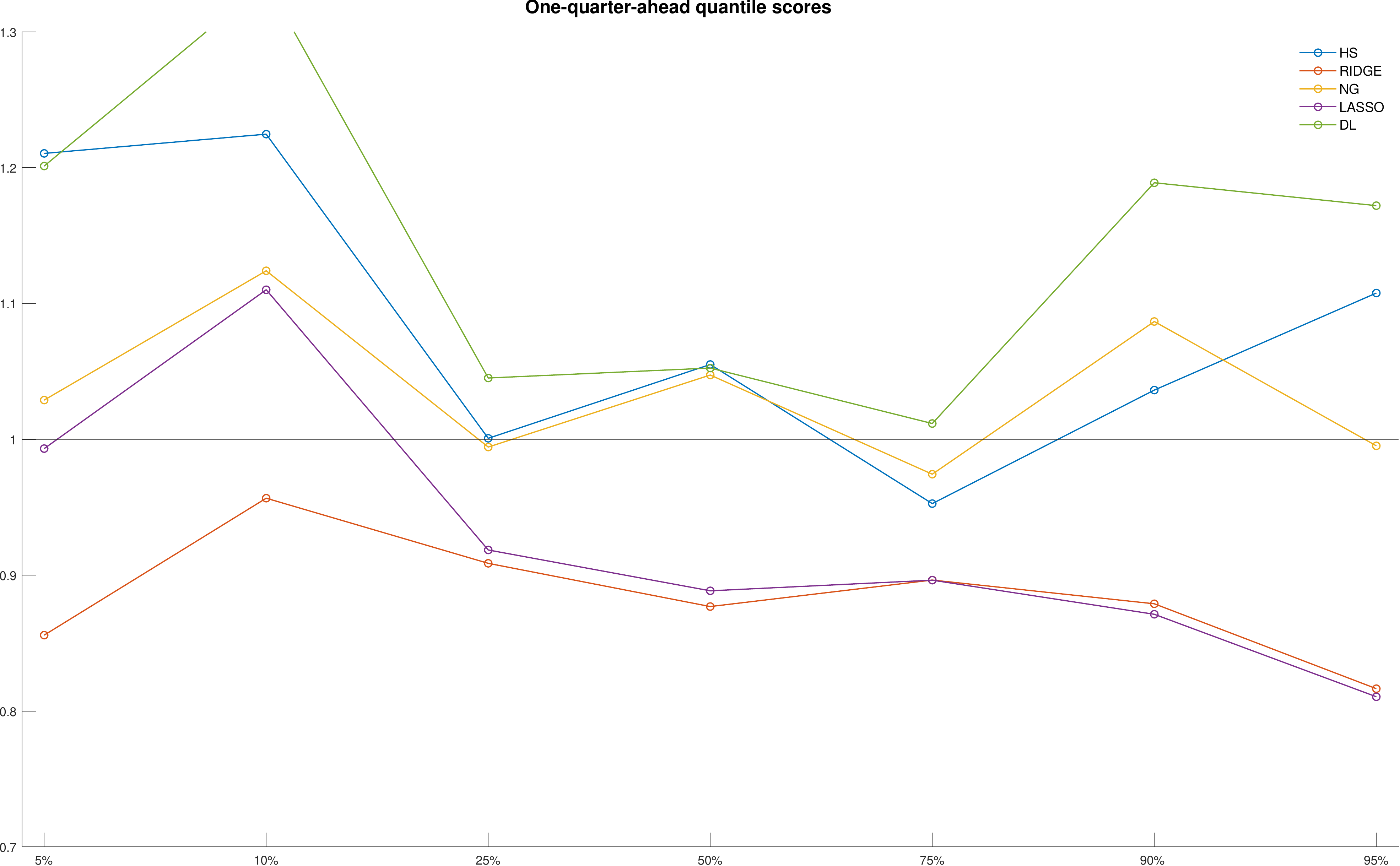} 
\caption[]{One-quarter-ahead quantile scores for different values of $\mathfrak{q}$, averaged over the hold-out period.}
\label{fig:linear_quantileloss_hfore=1start=1_Revision} 
\end{figure}

\begin{figure}[t!]
\centering 
\includegraphics[trim = 0mm 0mm 0mm 8mm,width=1\textwidth]{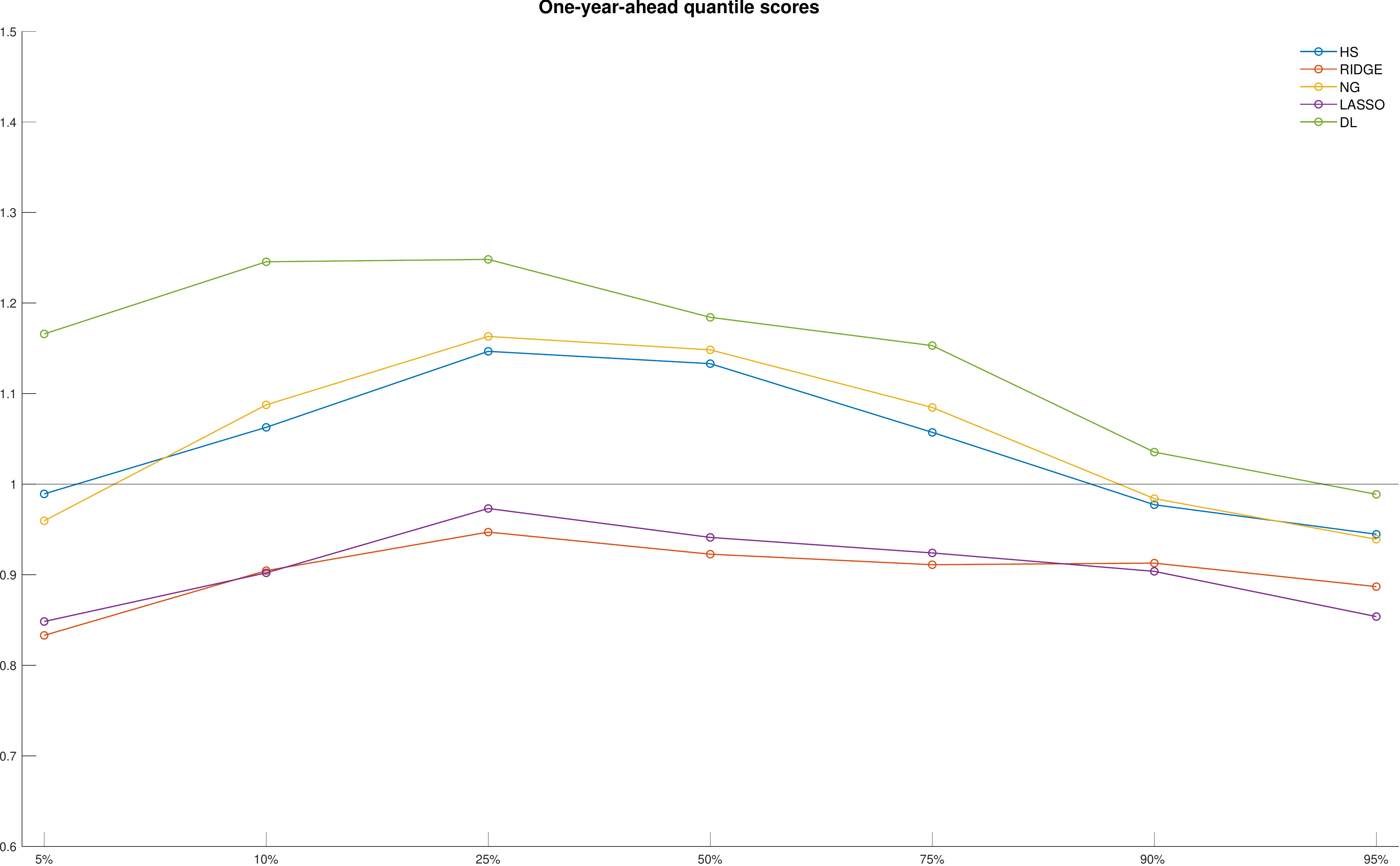} 
\caption[]{Four-quarters-ahead quantile scores for different values of $\mathfrak{q}$, averaged over the hold-out period.}
\label{fig:linear_quantileloss_hfore=4start=1_Revision} 
\end{figure}

Next, we drill deeper into the quantile-specific forecasting performance by considering QSs for $\mathfrak{q}$ ranging from $\mathfrak{q} \in \{0.05, 0.1, 0.25, 0.5, 0.75, 0.95, 0.99\}$. These, for one-step-ahead forecasts, are shown in Fig. \ref{fig:linear_quantileloss_hfore=1start=1_Revision} and Fig. \ref{fig:linear_quantileloss_hfore=4start=1_Revision} provides the four-steps-ahead results. Before starting our discussion, it is worth stressing that many of these differences to the benchmark model are statistically significant with respect to the DM test. The corresponding results are provided in the Online Appendix (see Figs. 15 and 16). Moreover, for the sake of brevity we leave out the results for Minn-Gamma from the graphs that follow. The reason is that they are indistinguishable from the ones of the Ridge. 

Similar to the findings based on the CRPSs, there is a great deal of heterogeneity across priors. Both the LASSO and in particular, the Ridge prior improve upon the benchmark for all quantiles by relatively large margins. These gains appear to be more pronounced in the tails, reaching up to 15-20 percent in terms of the QSs.  When focusing on the center of the distribution (i.e., the median forecast), the gains are a bit smaller. In general, the other priors perform considerably worse. The only exception turns out to be the NG prior, which displays an excellent performance in the left tail, while being still outperformed by the LASSO and the Ridge prior.

This brief discussion gives rise to a simple recommendation for practitioners. If interest is placed on producing precise tail forecasts (irrespective of the forecast horizon), it pays off to use large QRs coupled with either a LASSO, Minn-Gamma or Ridge-type prior. Since the Ridge prior is the simplest (i.e., it only features a single hyperparameter) and the empirical performance is very similar or better to the LASSO and (almost) identical to the Minn-Gamma, our focus from now on will be on comparing the Ridge-based QR with a range of non-linear specifications. 

\subsection{Allowing for nonlinearities in large scale QRs}\label{sec: nonlinearities}
In the previous subsection we have shown that using big QRs leads to tail forecasts that are superior to the ones of the benchmark ABG specification. Conditional on the quantile, these models are linear in the parameters. However, recent literature \citep[see, e.g.,][]{chkmp2021tail} suggests that nonlinearities become more important in the tails. Hence, we now address this question within our approximate framework. As discussed in Subsection \ref{sec: nonlinearities}, once we control for nonlinearities the number of covariates becomes huge. In this case, carrying out MCMC-based inference becomes impractical (i.e., for around 1,000 covariates estimating a model for a specific quantile takes over 15 minutes with MCMC and under a minute with VB). Hence, after having showed that VB-based forecasts are close to the ones obtained by estimating the models through MCMC, we focus on the comparison between linear QRs and nonlinear extensions, both estimated via VB-based methods.  

Table \ref{CRPSnonlinear} shows relative CRPSs for the different nonlinear models. As opposed to Table \ref{CRPSlinear}, all results are now benchmarked against the QR with the Ridge prior. This allows us to directly measure the performance gains from introducing nonlinearities relative to setting $g_\mathfrak{q}(\bm x_t)=0$. Notice that the absence of gray shaded cells in the table indicates that the DM test does not point towards significant differences in forecast accuracy between the linear and the different nonlinear QRs. 

Despite this, a few interesting insights emerge from the table. First, many numbers in the table are close to unity and differences are not statistically significant from the best performing linear QR.\footnote{For the GP-QR specifications with Ridge prior we obtain p-values between 0.1 and 0.2 using the test in \cite{DieboldMariano1995} with adjustments proposed by \cite{Harvey1997}.} This indicates that once we include many predictors, additionally controlling for nonlinearities of different forms only yields small positive (and sometimes negative) gains in terms of tail forecasting accuracy. Second, the first finding strongly depends on the approximation techniques chosen. Among all three specifications, using GPs is superior to using either polynomials or B-Splines to approximate the unknown function $g_\mathfrak{q}$. Second, and focusing on GP-QR specifications, the specific prior chosen matters appreciably. Whereas the results for the conditionally linear models clearly suggest that the LASSO and Ridge priors are producing the most precise density forecasts. The results for the nonlinear models tell a slightly different story. We observe that the Ridge does well again but, for one-quarter-ahead tail forecasts, is outperformed by the HS. The LASSO, in contrast, is the weakest specification. Since the LASSO is known to overshrink significant signals \citep[see, e.g.,][]{brown2010inference}, it could be that it misses out important information arising from the GP-based basis functions. Third, and finally, if we consider four-quarters-ahead predictions, the QR coupled with a GP and a Ridge prior becomes the single best performing model again. 
\begin{table}[t!]
\caption{CRPSs for nonlinear models}
\label{CRPSnonlinear}
\begin{tabular}{lcccc c cccc} 
\toprule
&\multicolumn{4}{ c }{One-quarter-ahead}&&\multicolumn{4}{ c }{Four-quarters-ahead} \\
\cline{2-5} \cline{7-10}
Model   &CRPS&CRPS-t&CRPS-l&CRPS-r&&CRPS&CRPS-t&CRPS-l&CRPS-r\\ \toprule
\multicolumn{4}{l }{Polynomials }& \multicolumn{1}{l }{\textit{   }}&& \\ \midrule
HS&	$1.02$ & $1.00$ & $1.08$&\B{}$0.94$  && $0.96$ & $0.97$ & $1.00$&$0.92$\\
Ridge&	$0.98$ &$0.94$ &$1.01$&$0.92$  &&$0.96$ & $0.97$ & $1.00$&$0.92$\\
NG&	$1.03$ & $0.99$ & $1.07$& $0.96$ &&$0.95$ & $0.96$ & $1.00$&$0.90$\\
LASSO&	$1.05$ &$1.04$ & $1.08$&$1.01$  &&$1.02$ & $1.00$ &$0.99$&$1.04$\\
DL&	$1.07$ & $1.04$ & $1.15$&$0.98$  && $1.22$ & $1.20$ & $1.29$&$0.94$\\
Minn-Gamma&	$0.98$ &$0.94$ &$1.01$&$0.92$  &&$0.96$ & $0.97$ & $1.00$&$0.92$\\  \midrule 
\multicolumn{4}{l }{B-Splines  }& \multicolumn{1}{l }{\textit{   }}&& \\ \midrule
HS&	$1.13$ & $1.16$ & $1.18$&$1.09$  && $1.10$ & $1.14$ & $1.05$&$1.18$\\
Ridge&	$1.08$ &$1.08$ &$1.08$&$1.08$  &&$1.09$ & $1.13$ & $1.04$&$1.17$\\
NG&	$1.15$ & $1.17$ & $1.20$&$1.12 $  &&$1.13$ & $1.17$ & $1.07$&$1.22$\\
LASSO&	$1.07$ &$1.06$ & $1.09$&$1.05$  &&$1.02$ & $1.01$ &$0.99$&$1.05$\\
DL&	$0.98$ & $1.00$ & $1.01$&$0.95$  && $1.04$ & $1.08$ & $1.02$&$1.09$\\ 
Minn-Gamma&	$1.08$ &$1.08$ &$1.08$&$1.08$  &&$1.09$ & $1.13$ & $1.04$&$1.17$\\ \midrule 
\multicolumn{4}{l }{Gaussian Processes  }& \multicolumn{1}{l }{\textit{   }}&& \\ \midrule
HS&	$0.96$ & $0.94$ & $0.97$& $0.94$   && $1.05$ & $1.02$ & $1.10$&$0.99$\\
Ridge&	$0.97$ &$0.95$ &$0.98$&$0.94$  &&$0.96$ & $0.95$ & $0.97$&$0.94$\\
NG&	$0.98$ & $0.95$ & $0.98$& $0.97$ &&$1.06$ & $1.02$ & $1.08$&$1.01$\\
LASSO&	$1.04$ &$1.04$ & $1.07$&$1.01$  &&$1.01$ & $0.99$ &$1.00$&$1.00$\\
DL&	$1.02$ & $0.97$ & $1.00$&$1.02$  && $1.22$ & $1.20$ & $1.29$&$1.10$\\ 
Minn-Gamma&	$0.97$ &$0.95$ &$0.98$&$0.94$  &&$0.96$ & $0.95$ & $0.97$&$0.94$\\ \midrule 
\end{tabular} 
\begin{tabular}{p{13cm}}
\footnotesize  \textit{Notes}: We highlight in light gray (dark gray) rejection of equal forecasting accuracy against the benchmark model at significance level 10\% (5\%) using the test in \cite{DieboldMariano1995} with adjustments proposed by \cite{Harvey1997}. Results are shown relative to the linear QR with a Ridge prior and are based on the full sample.
\end{tabular}
 \end{table}

\begin{figure}[h!]
\centering 
\includegraphics[trim = 0mm 0mm 0mm 00mm,width=1.05\textwidth]{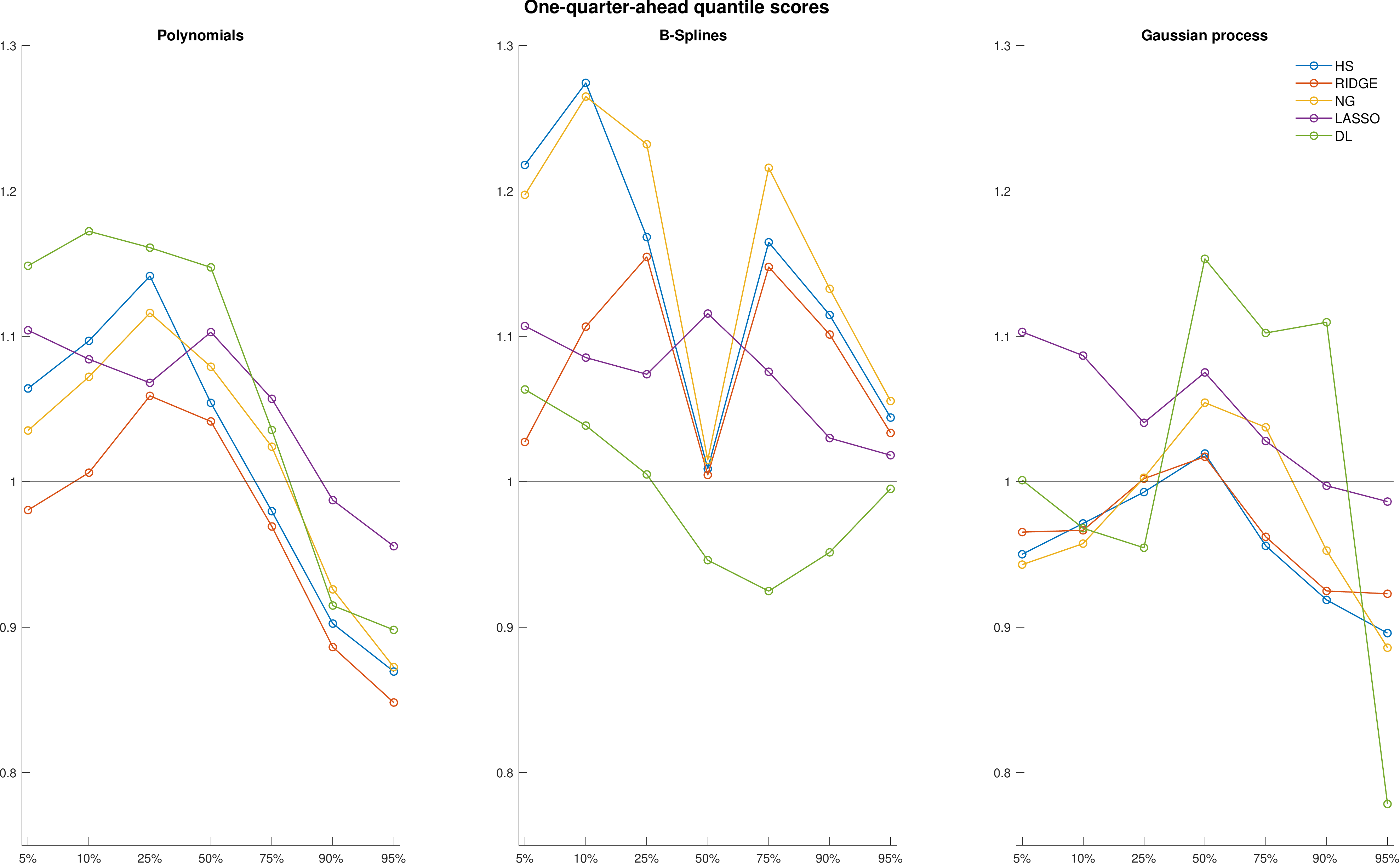} 
\caption[]{One-quarter-ahead quantile scores for different values of $\mathfrak{q}$, averaged over the hold-out period and normalized to the QR with a Ridge prior.}
\label{fig:non-linear_quantileloss_hfore=1start=1_Revision} 
\end{figure}

\begin{figure}[h!]
\centering 
\includegraphics[trim = 0mm 0mm 0mm 00mm,width=1.05\textwidth]{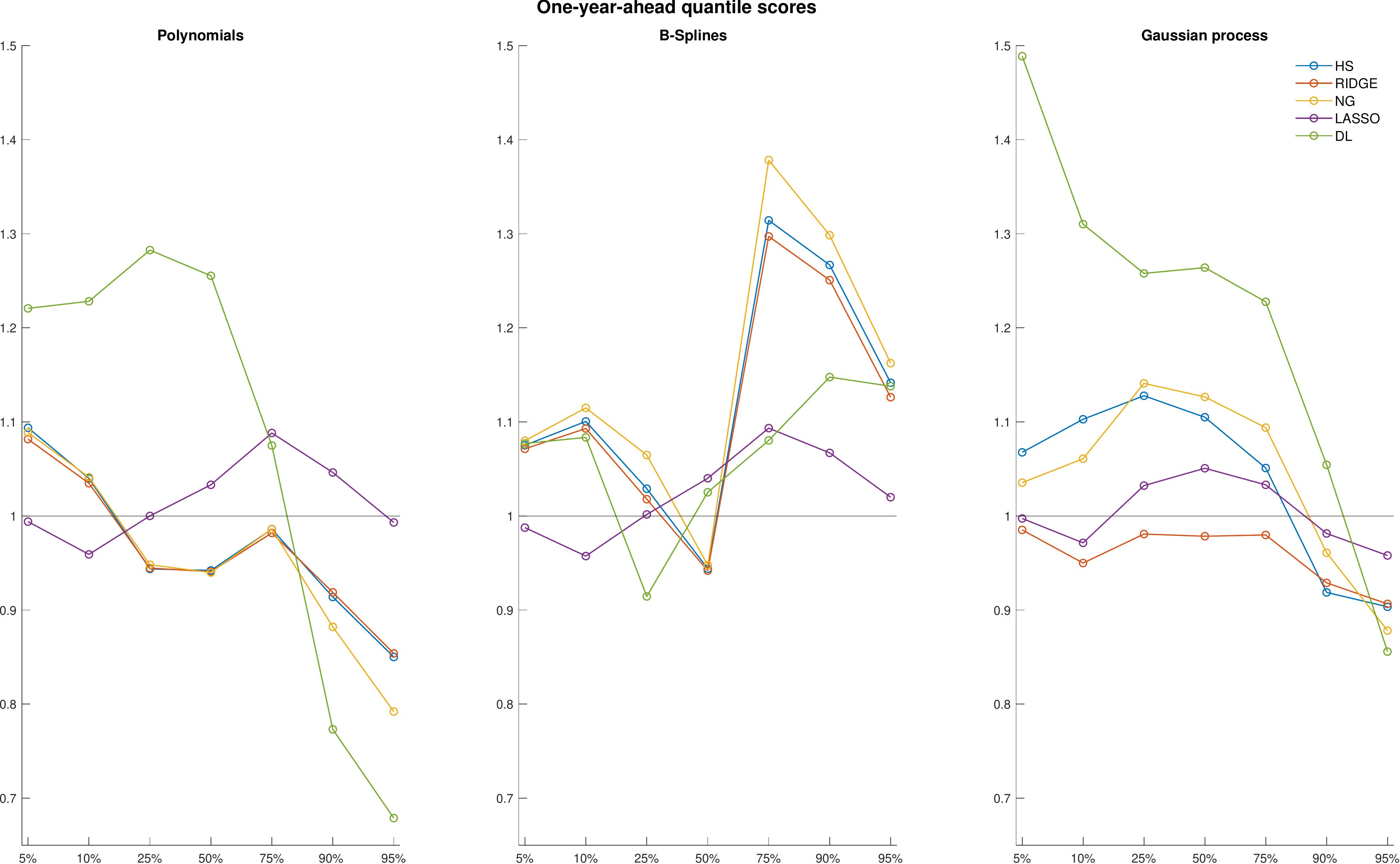} 
\caption[]{Four-quarters-ahead quantile scores for different values of $\mathfrak{q}$, averaged over the hold-out period and normalized to the QR with a Ridge prior.}
\label{fig:non-linear_quantileloss_hfore=4start=1_Revision} 
\end{figure}

To again gain a better understanding on which quantiles of the predictive distribution drive the CRPSs, Figs. \ref{fig:non-linear_quantileloss_hfore=1start=1_Revision} and \ref{fig:non-linear_quantileloss_hfore=4start=1_Revision} (similar to Figs. \ref{fig:linear_quantileloss_hfore=1start=1_Revision} and \ref{fig:linear_quantileloss_hfore=4start=1_Revision}) show the QSs for different quantiles. These are normalized to the linear QR with Ridge prior so that numbers smaller than one indicate that nonlinearities improve predictive accuracy for a given quantile and numbers exceeding one imply that nonlinearities decrease forecasting accuracy.

In general, both figures tell a consistent story: nonlinearities help in the right tail across both forecast horizons, for all three nonlinear specifications, and for most priors considered. The only exception to this pattern are four-quarters-ahead right tail forecasts of GDP growth when B-Splines are used. When there are gains, they are often sizable. For instance, in the case of the QR-GP model we observe accuracy improvements up to 25 percent relative to the linear QR model.

{}

When we focus on the left tail, accuracy premia often turn negative. In some  cases (such as for GP models with Ridge, NG and HS priors), there are accuracy gains for predicting downside risks but these gains are only rather small (reaching five percent in the case of the QR-GP regression with a Ridge prior). 

To sum up this discussion, our results indicate that linear models that exploit large datasets are sufficient if interest lays in capturing the left tail. In the right tail, it pays off to use nonlinear models. In Subsection \ref{sec: properties_fcst}, we investigate the properties of left and right-tail forecasts in order to shed some light on why the model is doing well in terms of predicting strong rebounds in economic activity.

\subsection{Heterogeneity of forecast accuracy over time} 
\label{ssub:heterogeneity_of_forecast_accuracy_over_time}
Up to this point, our analysis focused on averages over time. In the next step we will consider how the forecasting performance changes over the hold-out period. The overall results point towards limited gains of nonlinearities when the full hold-out period is considered. As a starting point, we ask whether nonlinear models improve tail forecasts if we focus only on the second half of the sample ($2006$Q2 to $2021$Q3). This period includes the GFC
 and the substantial downturn in output during the Covid-19 pandemic.

\begin{table}[h!]
\caption{CRPSs for nonlinear models: $2006$Q2 to $2021$Q3.}
\label{CRPSnonlinear_halfsample}
\begin{tabular}{lcccc c cccc} 
\toprule
&\multicolumn{4}{ c }{One-quarter-ahead}&&\multicolumn{4}{ c }{Four-quarters-ahead} \\
\cline{2-5} \cline{7-10}
Model   &CRPS&CRPS-t&CRPS-l&CRPS-r&&CRPS&CRPS-t&CRPS-l&CRPS-r\\ \toprule
\multicolumn{4}{l }{Polynomials }& \multicolumn{1}{l }{\textit{   }}&& \\ \midrule
HS& $0.97$ & $0.93$ & $0.97$&$0.86$  && $0.88$ & $0.86$ & $0.94$&$0.80$\\
Ridge&  $0.91$ &$0.86$ &$0.96$&$0.83$  &&$0.88$ & $0.87$ & $0.94$&$0.81$\\
NG& $0.98$ & $0.92$ & $1.01$& $0.89$ &&$0.87$ & $0.84$ & $0.94$&$0.76$\\
LASSO&  $1.03$ &$1.02$ & $1.01$&$0.99$  &&$0.98$ & $0.95$ &$0.95$&$0.99$\\
DL& $1.01$ & $1.05$ & $1.02$&$0.98$  && $1.02$ & $0.97$ & $1.05$&$0.86$\\
Minn-Gamma&  $0.91$ &$0.86$ &$0.96$&$0.83$  &&$0.88$ & $0.87$ & $0.94$&$0.81$\\  \midrule 
\multicolumn{4}{l }{B-Splines  }& \multicolumn{1}{l }{\textit{   }}&& \\ \midrule
HS& $1.02$ & $1.02$ & $1.08$&$0.95$  && $0.94$ & $0.93$ & \B{}$0.88$&$0.99$\\
Ridge&  $0.96$ &$0.92$ &$0.95$&$0.94$  &&$0.93$ & $0.92$ & \B{}$0.88$&$0.99$\\
NG& $1.02$ & $1.01$ & $1.07$&$0.97 $  &&$0.95$ & $0.94$ & $0.89$&$1.01$\\
LASSO&  $1.07$ &$1.05$ & $1.09$&$1.02$  &&$0.99$ & $0.97$ &$0.95$&$1.01$\\
DL& $0.90$ & $0.88$ & $0.93$&$0.85$  && $0.96$ & $0.93$ & $0.94$&$0.96$\\ 
Minn-Gamma&  $0.96$ &$0.92$ &$0.95$&$0.94$  &&$0.93$ & $0.92$ & \B{}$0.88$&$0.99$\\ \midrule 
\multicolumn{4}{l }{Gaussian Processes  }& \multicolumn{1}{l }{\textit{   }}&& \\ \midrule
HS& $0.94$ & $0.91$ & $0.96$& $0.91$   && $0.97$ & $0.96$ & $1.02$&$0.91$\\
Ridge&  \B{}$0.95$ &\B{}$0.93$ &\B{}$0.97$&\B{}$0.92$  &&$0.94$ & $0.92$ & \B{}$0.95$&$0.91$\\
NG& $0.97$ & $0.93$ & $0.97$& $0.95$ &&$0.99$ & $0.96$ & $1.01$&$0.94$\\
LASSO&  $1.06$ &$1.05$ & $1.09$&$1.01$  &&$1.00$ & $0.98$ &$0.99$&$0.99$\\
DL& $1.02$ & $0.96$ & $0.99$&$1.02$  && $1.05$ & $1.04$ & $1.03$&$0.96$\\ 
Minn-Gamma&  \B{}$0.95$ &\B{}$0.93$ &\B{}$0.97$&\B{}$0.92$  &&$0.94$ & $0.92$ & \B{}$0.95$&$0.91$\\ \midrule 
\end{tabular} 
\begin{tabular}{p{13cm}}
\footnotesize  \textit{Notes}: We highlight in light gray (dark gray) rejection of equal forecasting accuracy against the benchmark model at significance level 10\% (5\%) using the test in \cite{DieboldMariano1995} with adjustments proposed by \cite{Harvey1997}. Results are shown relative to the linear QR with a Ridge prior.
\end{tabular}
 \end{table}

Table \ref{CRPSnonlinear_halfsample} is the same as Table \ref{CRPSnonlinear} but includes only averages of CRPSs over the $2006$Q2 to $2021$Q3 period. The table reveals much larger gains from using nonlinear models if the volatile part of the sample is used as our verification period. In particular, we find significant improvements of up to seven percent in  the tails for the GP-QR-Ridge model. When we focus on QR with polynomials, we find even stronger gains (which reach around 17 percent for the right tail) but these are never significant. There are also some nonlinear specifications that improve upon the linear benchmark when the overall CRPS is used (such as GP-QR-Ridge). 

When we consider  higher order predictions, gains become even larger, corroborating the findings in \cite{chkmp2021tail}. For longer horizons, spline-based models coupled with a HS or Ridge prior yield one-year-ahead left tail forecasts which are around 12 percent more accurate than the ones of the linear QR model with the Ridge prior.  

\begin{figure}[h!]
\centering 
\includegraphics[trim = 0mm 0mm 0mm 00mm,width=1.05\textwidth]{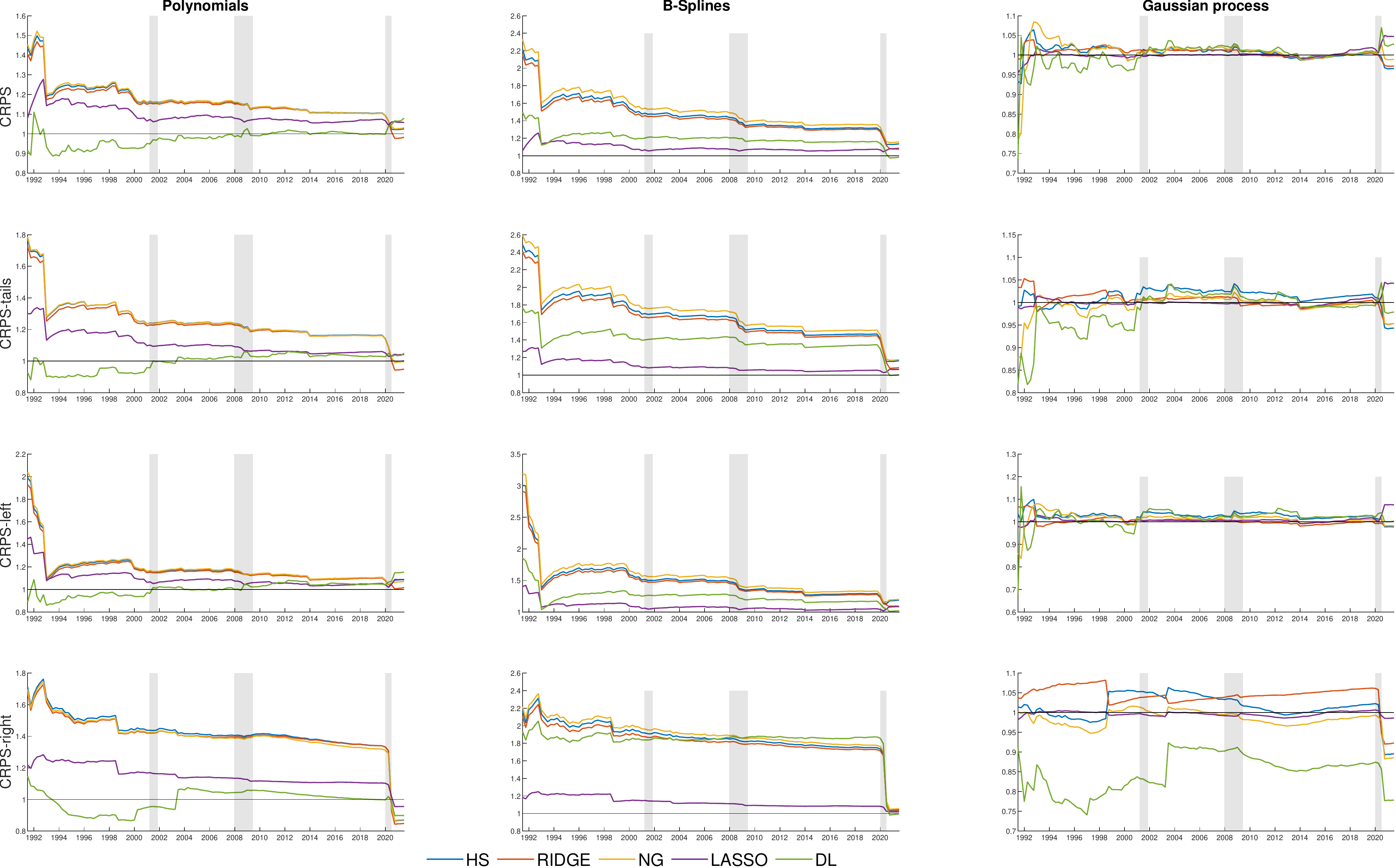} 
\caption[]{Cumulative one-quarter-ahead CRPS relative to the linear QR with the Ridge prior over the hold-out period.}
\label{fig:non-linear_cumsumCRPSnhfore=1_Revision} 
\end{figure}

\begin{figure}[h!]
\centering 
\includegraphics[trim = 0mm 0mm 0mm 00mm,width=1.05\textwidth]{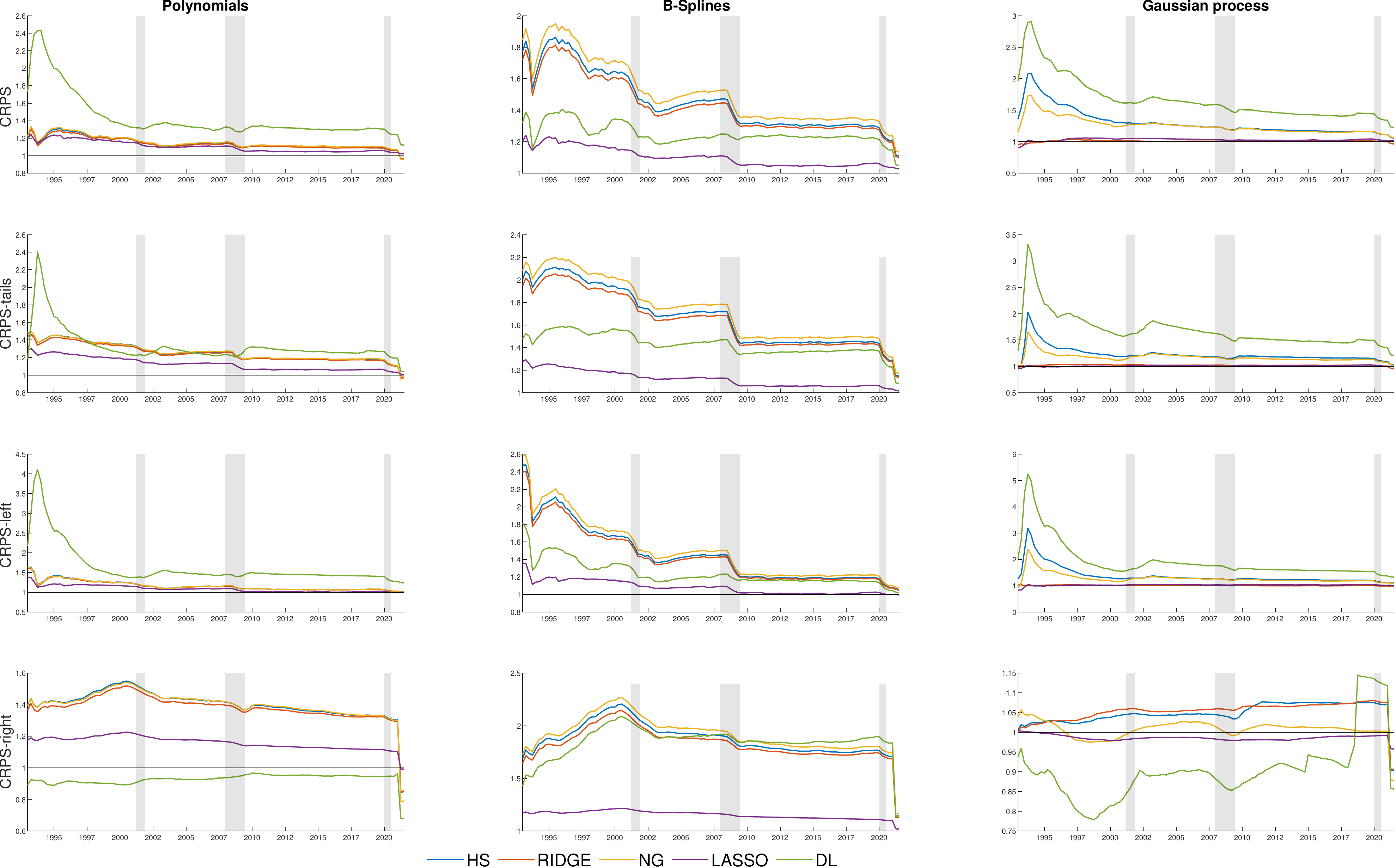} 
\caption[]{Cumulative four-quarter-ahead CRPS relative to  linear QR with a Ridge prior over the hold-out period.}
\label{fig:non-linear_cumsumCRPSnhfore=4_Revision} 
\end{figure}

This discussion shows that if we focus on periods characterized by a rather volatile macro environment, nonlinearities appear to be important. In the next step, we investigate whether a split of the hold-out sample in pre-2006 and post-2006 still masks important temporal idiosyncrasies. To focus on the time-evolution of the CRPSs,  Figs. \ref{fig:non-linear_cumsumCRPSnhfore=1_Revision} and  \ref{fig:non-linear_cumsumCRPSnhfore=4_Revision}  show the cumulative CRPSs relative to the linear benchmark QR (with a Ridge prior) for one-quarter and four-quarters-ahead forecasts.

We start by focusing on the one-quarter-ahead forecasts. For this specification, the density forecasting performance  is heterogenous over time. In the first part of the sample, models using either polynomials or Gaussian processes coupled with a DL prior yield CRPSs that are superior to the linear benchmark. However, these accuracy gains vanish during the GFC. When we put more weight on tail forecasting accuracy (and consider GP-QRs), the gains disappear as early as during the 2001 recession that followed the 9/11 terrorist attacks and the burst of the dot-com bubble.  

In the pandemic, we observe a sharp increase in predictive accuracy for several priors (most notably the Ridge and NG priors). This pattern is more pronounced for the weighted variants of the CRPSs. Considering the other nonlinear model specifications gives rise to similar insights. Spline-based approximations to $g_\mathfrak{q}$ generally perform poorly up until the pandemic. During the pandemic, even this specification improves sharply against the linear benchmark specification. This pattern is particularly pronounced for the GP-QRs.

Considering the performance of the models and priors that did well on average (GP-QRs with Ridge and the HS) reveals that most of these gains are actually driven by a superior performance during turbulent times. Specifically, gains arise from a superior performance during the GFC and the Covid-19 pandemic, as evidenced by declining ratios of CRPSs over these periods. 

Turning to four-quarters-ahead forecasts provides little new insights. Models using the DL prior do not excel in the first part of the hold-out period and are generally outperformed by the linear QR. However, accuracy improvements during the GFC and the pandemic are quite pronounced for splines and the GP-QRs. This pattern is even more pronounced than in the case of the one-quarter-ahead forecasts.

To sum up this discussion, our results indicate that forecast performance is heterogenous over time. Different models such as the Polynomial-QR and the GP-QR with a DL prior outperform in the early part of the hold-out period. This performance premium vanishes during the first two recessions observed in the sample.  In contrast, other models such as QR-GP with either the NG or the LASSO do not gain much in tranquil periods but excel during recessions.

\subsection{Properties and determinants of the quantile forecasts} \label{sec: properties_fcst}
The previous sub-sections have outlined that QRs and QRs with nonlinear components perform well in terms of tail forecasting. In this sub-section, our goal is to investigate which variables determine the quantile forecasts and in what respect successful shrinkage priors differ from their less successful counterparts.

The presence of nonlinearities complicates our investigation since it is not clear how to measure the effect of $\bm x_t$ on a given quantile of $y_t$ in the presence of nonlinearities. Moreover, given that $K$ is large, it is difficult to understand which variables have an important effect on the quantile forecast of $y_t$. This issue is further intensified since priors such as the Ridge imply that most elements in $\bm x_t$ have a small effect and thus the importance of a single variable is difficult to quantify. As a simple solution to both issues, we follow \cite{clark2022forecasting} and approximate the nonlinear, quantile-specific model using a linear posterior summary \citep[see][]{woody2021model}. Specifically, we estimate the following regression model:
\begin{equation*}
     Q_{\mathfrak{q}, t} = \bm x'_t \hat{\bm \alpha}_\mathfrak{q} + \hat{\varepsilon}_t,\quad \hat{\varepsilon}_t \sim \mathcal{N}(0, \sigma_{\hat{\varepsilon} \mathfrak{q}}^2).
 \end{equation*} 
 On the linearized coefficients we use a Horseshoe prior and on the error variances an inverse Gamma prior. To achieve interpretability and decouple shrinkage and selection \citep[see][]{hahn2015decoupling}, we then apply the SAVS estimator proposed in \cite{ray2018signal} to the posterior mean of $\hat{\bm \alpha}_\mathfrak{q}$.\footnote{\cite{huber2021inducing} and \cite{hauzenberger2021combining} apply SAVS to multivariate time series models and show that it works well for forecasting.} This will yield a sparse variant of $\hat{\bm \alpha}_\mathfrak{q}$ that is easy to interpret and can be understood as the best linear approximation to the corresponding quantile forecast arising from the nonlinear model.  We label predictors that survive the sparsification process as robust. Notice that this does not imply that the corresponding forecasting model is sparse. It simply tells us that, conditional on a given shrinkage prior, a given predictor plays an important role in approximating a given quantile.

 For brevity, we focus on one-step-ahead forecasts. Results for four-quarters-ahead are included in the Online Appendix.

\begin{figure}[h]
\centering 
\includegraphics[trim = 0mm 0mm 0mm 00mm,width=1\textwidth]{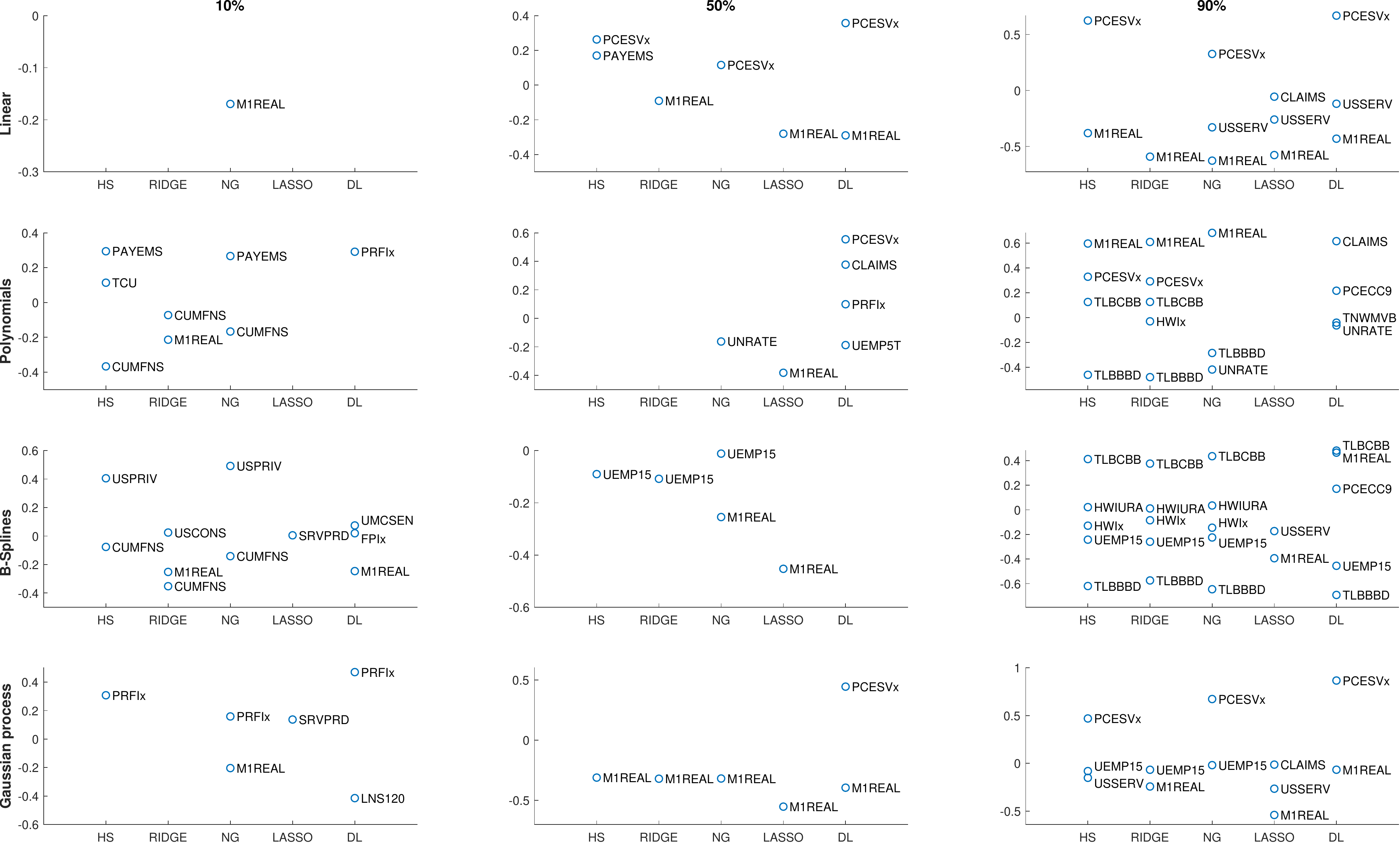} \caption[]{One-quarter-ahead linearized posterior summaries across quantiles}
\label{fig:Boxplots_forecastsnhfore=1Covid=1_Revision} 
\end{figure}

Figure \ref{fig:Boxplots_forecastsnhfore=1Covid=1_Revision} shows the results of this exercise across nonlinear specifications and priors. Starting with the left tail forecasts and linear models suggests that most quantile forecasts are not related to elements in $\bm x_t$ in a robust manner. There is only one exception. In the case of the QR with the NG prior, we find that real money growth (M1real) survives the sparsification step and the relationship indicates that declines in money growth imply an increase in tail risks (i.e. a decline in GDP growth in the ten percent quantile). 

If we focus on nonlinear models other variables appear to be correlated with forecasts of tail risks. Among the different priors, we find some  variables which show up repeatedly. Among these are all nonfarm employees (PAYEMS), money growth, capacity utilization in manufacturing (CUMFNS) and private fixed investment (both  residential and non-residential). Most of these variables are forward looking in nature and thus consistent with our intuition that economic agents form expectations about the state of the economy in the future and thus change their investment decisions accordingly. Notice that the relationship between private fixed investment is particularly pronounced for GP-QRs under the HS, NG and the DL prior. Another pattern worth mentioning is that the LASSO-based forecasts are generated from sparse models across both linear and nonlinear specifications.

Once we focus on the center of the distribution, we find that forecasts from linear models are driven by one or two variables.  Most prominently, specifications that do well in terms of point forecasts (such as the Ridge and LASSO) yield point forecasts that display a strong relationship with (lagged) money growth. In case we adopt a nonlinear specification, some differences arise across specifications. For polynomials, median forecasts under all priors except the DL are related to very few predictors, with money growth and short-run unemployment showing up for the NG and LASSO models. The DL prior implies a more dense model. This could be a possible reason for the rather weak performance of this specification. When we turn to spline-based models, we again find a similar pattern. Money growth shows up in the case of the NG and LASSO and short-run unemployment predicts median output growth if we adopt a HS, Ridge or NG prior. Models that capture nonlinearities through GPs, our best performing nonlinear specifications, give rise to a very consistent pattern across priors. In all cases, lagged money growth appears to be a robust predictor of GDP growth and  it impacts GDP growth forecasts negatively.

Finally, when our focus is on right-tail forecasts, all models become much more dense. Variables that have been showing up in the case of left-tail and point forecasts show up again (most notably money growth and short-term unemployment). Additional variables such as initial unemployment claims or prices remain in the sparse model as well. But there is no clear pattern across models, except for the fact that money growth also remains in the set of robust predictors even if much shrinkage is introduced. In general, we can say that as opposed to median forecasts, right tail predictions are driven by more variables. This pattern also holds (for some nonlinear specifications) for left tail forecasts but to a lesser degree.

The analysis based on linearized coefficients provides information on which variables are predictive for output growth forecasts across quantiles. However, the analysis in Subsections \ref{sec: linear} and \ref{sec: nonlinear} suggests that differences in forecast performance are driven by the prior. To understand which properties of a given prior exert a positive effect on predictive accuracy, we now focus on the shrinkage hyperparameters of the different priors. Comparing the amount of shrinkage introduced through the different priors is not straightforward. Here, our measure of choice is based on using the re-scaled log determinant of the prior covariance matrices as a measure of overall shrinkage for each respective prior. Since all prior covariance matrices are diagonal, this simply amounts to summing over the log of the diagonal elements of $\bm B_{0 \mathfrak{q}}$ and then normalizing through by the number of diagonal elements. This constitutes a rough measure of overall shrinkage and we can compute it for each quarter in the hold-out period. Again, we will focus on shrinkage introduced in one-quarter-ahead predictive regressions. The four-quarters-ahead results are qualitatively similar and included in the Online Appendix.

\begin{figure}[h]
\centering 
\includegraphics[trim = 0mm 0mm 0mm 00mm,width=1\textwidth]{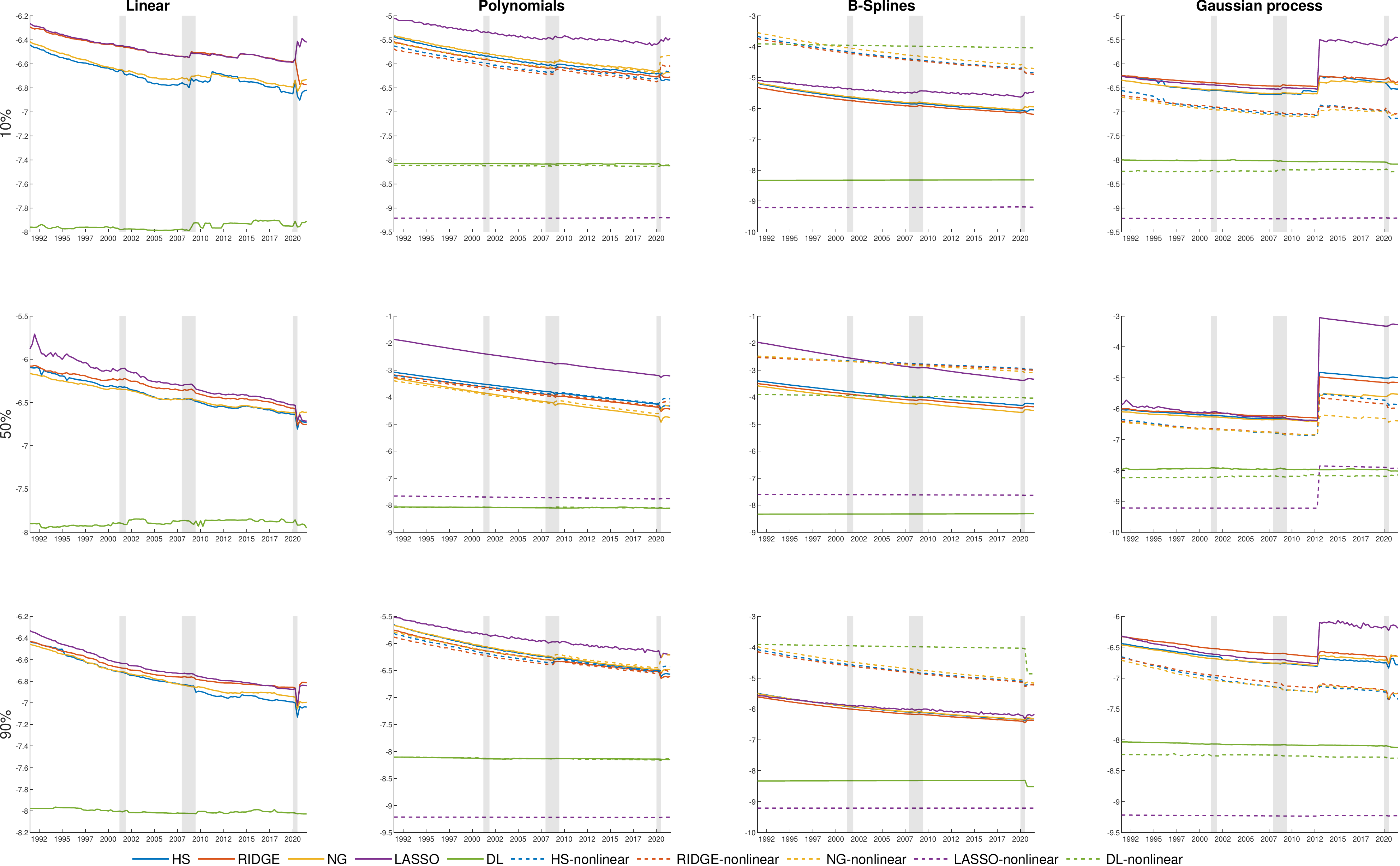} 
\caption[]{Overall shrinkage in one-step-ahead predictive QRs}
\label{fig:Hyperparameter_localnhfore=1_Revision} 
\end{figure}

Log-determinants of the prior covariance matrices over the hold-out period are depicted in Fig. \ref{fig:Hyperparameter_localnhfore=1_Revision}. The figure includes (if applicable) solid lines which refer to the amount of shrinkage introduced on the linear coefficients and dashed lines which refer to the log-determinants of the prior covariances that relate to the shrinkage factors on the basis coefficients of the different nonlinear models.

From this figure, a few interesting insights emerge.  First, the different priors introduce different degrees of shrinkage. Overall, two priors stand out in terms of the amount of shrinkage they introduce. The first one is the DL. This is rather surprising given the fact that this prior performs worst in the forecasting horse race but also leads to posterior summaries which feature several non-zero coefficients. Our conjecture is that this prior forces the vast majority of coefficients to effectively zero but several coefficients remain sizable and the corresponding set of variables is still too large causing overfitting issues to arise. The prior that introduces the largest amount of shrinkage is the LASSO. In this case, almost all coefficients are very small. These observations are corroborated by boxplots, included in the Online Appendix (see Figs. 3 to 6 in the Online Appendix), which show the scaling parameters over three sub-samples. Our results imply that models which feature a large number of shrunk coefficients provide better forecasts than models featuring many coefficients that are effectively zero and some coefficients that are non-zero and sizeable. This is consistent with findings in \cite{giannone2021economic}, who provide empirical evidence that macroeconomic data is rather dense as opposed to sparse. Notice that the fact that dense models produce accurate tail forecasts is not  inconsistent with our analysis based on linearized posterior summaries. This is because the linearized model under a shrinkage and sparsification approach strikes a balance between achieving a good model fit while keeping the model as simple as possible. Hence, if the covariates in the panel co-move,  shrinkage and sparsification techniques  will select one of these variables.

Second, in almost all cases, the amount of shrinkage introduced on the nonlinear part of the different models is much larger than the degree of shrinkage on linear coefficients. This holds for most priors, nonlinear methods and over all time periods. One exception is the Spline-QR specification with a DL prior and when the right tail is considered. Interestingly, this specific combination of much stronger shrinkage on the linear part of the model and less shrinkage on the nonlinear part leads to good forecasts in the right tail (see Fig. \ref{fig:non-linear_quantileloss_hfore=1start=1_Revision}).

Third, and finally, there is (with some notable exceptions) relatively little time-variation in the amount of shrinkage over the hold-out period. The only exception are the GP-QRs. In this case, the amount of shrinkage decreases appreciably from 2013 onward. 


\section{Other applications of our VB-based QR estimator} \label{extensions}
As repeatedly stated in the paper, our approach is highly scalable and the general framework can be applied to estimate specifications which are huge dimensional. In this section, we sketch how our approach can be used to estimate two popular models in the macro forecasting literature. 

When predicting GDP growth, it might pay off to exploit information from higher frequency sources. This can be achieved through mixed frequency models such as the U-MIDAS \citep{ghysels2007midas,foroni2015unrestricted}. In this case,  $g_\mathfrak{q}(\bm x_t)$ can be replaced with a linear function $g_\mathfrak{q}(\bm s_t)$, where $\bm s_t$ contains the quarterly observations of the higher frequency variables where the different elements in $\bm s_t$ relate to different periods (days, weeks, months) within a quarter. As an example, suppose that $\bm s_t$ consists of $N$ different weekly series and a quarter consists of 13 weeks. In this case, $M = 13 N$ and if $N$ is large, the number of parameters becomes huge.  If the researcher is interested in using high frequency information without introducing parametric restrictions, our VB-based approach could handle this situation particularly well. 

Another prominent example which we mentioned in Subsection \ref{sec: nonlinearities} are time-varying parameter QRs \citep{korobilis2021,pfarrhofer2022modeling}. In this case, we set $\bm g_\mathfrak{q} = \bm Z \bm \gamma_\mathfrak{q}$ where $\bm Z$ is a lower triangular matrix given by:
\begin{equation*}
   \bm Z = \begin{pmatrix}
     \bm x'_1 & 0 & \dots & 0 \\
     \bm x'_2 & \bm x'_2 & \dots & 0 \\
     \vdots & \vdots & \ddots & 0\\
     \bm x'_T & \bm x'_T & \dots & \bm x'_T
   \end{pmatrix},
 \end{equation*}
 which is a $T \times TK$ matrix and $\bm \gamma_\mathfrak{q}$ is a $TK-$dimensional vector of regression coefficients. In this case, the model is a QR with time-varying parameters that evolve according to a random walk. Since $K$ and $T$ can be large, the dimension of the problem quickly becomes intractable. Again, our approach could handle such a situation well and several recent papers advocate using VB to carry out estimation and inference in TVP regression models \citep[see][]{koop2020bayesian}. By contrast, MCMC estimation of such models is possible but requires sophisticated tricks or other approximations to speed up computation \citep[for a recent contribution that uses singular value decompositions, see][]{hauzenberger2022fast}.

\section{Conclusion}
In this paper, we have shown that combining QRs with nonlinear specifications and large datasets leads to precise quantile forecasts of GDP growth. Since the resulting models are high dimensional, we consider several popular shrinkage priors to regularize estimates. MCMC-based estimation of these huge dimensional models is slow. Hence, we speed up computation by using VB approximation methods that approximate the joint posterior distribution using simpler approximating densities.

The empirical results indicate that our methods work remarkably well when the CRPS is taken under consideration. When we put more weight on the tail forecasting performance, we find that most of the overall gains are driven by a strong performance in both the left and right tail while the performance in the center of the distribution is close to the predictive accuracy of the simple quantile regression proposed in \cite{adrian2019vulnerable}. These results, however, differ across priors and nonlinear specifications. In principle, it can be said that models featuring simple shrinkage priors, such as the LASSO or Ridge, in combination with GPs to capture nonlinearities of arbitrary form yield the most precise forecasts. 

%

\bibliographystyle{apalike}
\bibliography{lit}
\appendix

\renewcommand{\thesection}{Online Appendix \Alph{section}}
\setcounter{table}{0}
\renewcommand{\thetable}{\Alph{section}.\arabic{table}}
\setcounter{figure}{0}
\renewcommand{\thefigure}{\Alph{section}.\arabic{figure}}

\end{document}